\documentclass[aps,prx,twocolumn,superscriptaddress,longbibliography]{revtex4-1}

\usepackage[T1]{fontenc}
\usepackage[utf8]{inputenc}
\usepackage{lmodern}
\usepackage[sumlimits]{amsmath} 
\usepackage{amssymb}
\usepackage{yhmath}
\usepackage{graphicx}
\usepackage{epstopdf}
\usepackage{latexsym}
\usepackage{color}
\usepackage{nicefrac}
\usepackage{dsfont} 
\usepackage{wasysym} 
\usepackage{stmaryrd}
\usepackage{hyperref} 
\usepackage{xcolor}
\usepackage{multirow}
\usepackage{soul}
\usepackage{cancel}

\usepackage{esint}
\usepackage{tikz}

\definecolor{airforceblue}{rgb}{0.36, 0.54, 0.66}

\newcommand{\be}{\begin{equation}}
\newcommand{\ee}{\end{equation}}
\newcommand{\bea}{\begin{eqnarray}}
\newcommand{\eea}{\end{eqnarray}}

\usepackage{verbatim}
\usepackage{stmaryrd}
\usepackage[sumlimits]{amsmath} 
\usepackage{dsfont} 
\usepackage{caption}
\usepackage{cancel}
\usepackage{url}

\usepackage{subfigure}

\newcommand{\bs}{\boldsymbol}

\usepackage{scalerel}

\usepackage{accsupp}

\begin{document}

\title{Analytic Formulas for the Anomalous Hall Effect in Itinerant Magnets}

\author{Lucile Savary}
\affiliation{French American Center for Theoretical Science,
  CNRS, KITP, University of California, Santa Barbara, CA 93106-4030}
\affiliation{Kavli Institute for Theoretical Physics, University of
  California, Santa Barbara, CA 93106-4030}
\affiliation{\'{E}cole Normale Sup\'{e}rieure de
  Lyon, CNRS, Laboratoire de physique, 46, all\'{e}e
  d'Italie, 69007 Lyon}

\date{\today}
\begin{abstract}
We clarify the origin of what is sometimes called the ``topological
anomalous Hall effect,'' provide analytical formulas to compute all
the contributions to the Hall conductivity in the presence of
Kondo-coupled spins and spin orbit coupling. The derivation is technical but we emphasize that the
results can be very easily applied.
\end{abstract}

\maketitle

{\bf {\em Introduction.---}}The electrical Hall effect, whereby a current transverse
to an applied electric field can flow, proceeds from the coupling of a magnetic field 
to itinerant electrons. The (electrical) anomalous Hall effect
\cite{nagaosa2006,culcer2024} refers
to the same phenomenon when the Hall conductivity is not directly
proportional to an applied magnetic field and, without skew scattering, generally arises from an
electronic Berry curvature. Given the momentum-space Berry curvature for a band $n$
of (effectively)
free electrons at each
point in the Brillouin zone $\bs{k}$,
$\Omega_{(n)}^\gamma(\bs{k})=i\epsilon_{\alpha\beta\gamma}\partial_{k_\alpha}[\langle
u_n(\bs{k})|\partial_{k_\beta}u_n(\bs{k})\rangle]$, with
$\alpha,\beta,\gamma$ cartesian coordinates, $|u_n(\bs{k})\rangle$ the unit cell periodic part of the
single-electron wavefunction in band $n$, the TKNN formula provides 
the resulting electrical Hall conductivity,
$\sigma^{xy}_{\rm
  H}=\sum_n\int_{\bs{k}}\Omega_{(n)}^z(\bs{k})f_n(\bs{k})$, 
where $f_n(\bs{k})$ is the electronic filling function of band $n$ at momentum
$\bs{k}$. In turn, ``sources''
of Berry flux such as band crossings, e.g.\ in Weyl semimetals, will
then naturally produce a nonzero Hall conductivity \cite{burkov2011}. While the
Berry curvature appearing in the Hall conductivity formula
\cite{karplus1954,adams1959,thouless1982} written above may be a characteristic
of the pure (`intrinsic') electronic bands (we mean a band structure with spin-orbit coupling), it can also arise from a
`reconstructed' band structure resulting from the (`extrinsic')
coupling of the charge carriers to other degrees of freedom. This is {\em
  expected} to be the case in particular when electrons couple to
spins which locally (or globally) display nonzero spin chirality,
$\chi_{ijk}=\vec{S}_i\cdot(\vec{S}_j\times\vec{S}_k)$ for
spins at three site $i,j,k$ or when the spin-orbit structure and a
(possibly coplanar) spin structure conspire to produce a nonzero Berry
curvature
\cite{ye1999,chen2014,zhang2020,li2023}. $\chi_{ijk}$
can be viewed as a `real space Berry curvature' \cite{zhang2020} and
continuum versions such as
$\vec{n}\cdot(\partial_{x_\mu}\vec{n}\times\partial_{x_\nu}\vec{n})$
appear in semiclassical (long-wavelength) approaches \cite{wickles2013,mangeolle2024}. In practice the
effects of both the structure of the pure electronic bands and that of
the coupling to other degrees of freedom combine and couple.  To the
best of our knowledge, only within some approximations (e.g.\ double
exchange \cite{karplus1954}) and under
specific assumptions (e.g.\ long-wavelength limit) has the explicit relation between `real space Berry
curvature' and Hall conductivity been shown, and the
effects of spin-orbit coupling have only been analyzed numerically.

Here, we derive analytical formulas for the Hall conductivity directly
in terms of (i) real space structures such as spin chiralities {\em within a unit
cell} and (ii) spin-orbit coupled hopping terms, and include all combinations of
effects. To proceed, we make use of formulas for the Berry
curvature and quantum metric in terms of projection operators, and
expressions for the projectors as polynomials of the Hamiltonian
matrix. 

\begin{figure}[htbp]
  \centering
  \includegraphics[width=.8\columnwidth]{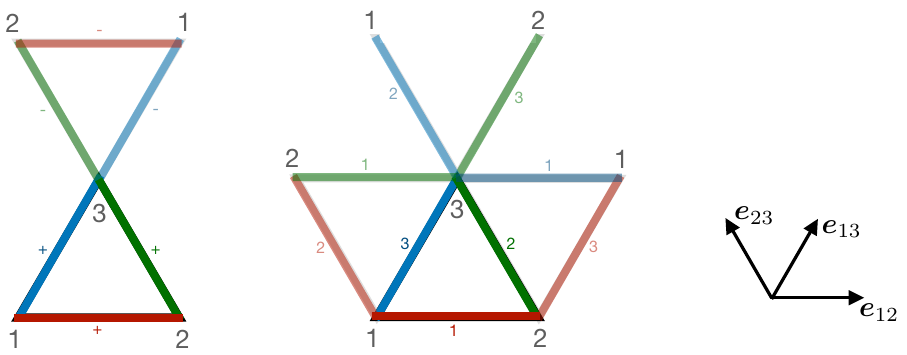}
  \caption{Three-sublattice cases: kagom\'e (left) and
    three-sublattice triangular (middle)
    lattices, with $\bs{e}_{12}^{(0)}=\bs{e}_{12}^{(1)}=(1,0)$,
    $\bs{e}_{23}^{(0)}=\bs{e}_{23}^{(2)}=\frac{1}{2}(-1,\sqrt{3})$, $\bs{e}_{31}^{(0)}=\bs{e}_{31}^{(3)}=\frac{-1}{2}(1,\sqrt{3})$.}
  \label{fig:three}
\end{figure}

{\bf {\em Kondo-coupled band structure.---}}While the formalism developed
here applies beyond this specific case, let us focus for concreteness on
the case of a Kondo-coupled band structure with
hamiltonian ${\rm H}=\frac{1}{\sqrt{N_{\rm u.c.}}}\sum_{\bs{k}}\Psi^\dagger_{\bs{k}}\hat{H}(\bs{k})
\Psi^{\vphantom{\dagger}}_{\bs{k}}$ where $N_{\rm u.c.}$ is the number
of unit cells, $\Psi^\dagger_{\bs{k}}$ is
an $M=2N$ vector of spin-1/2 fermion creation operators, where $N$ is
the number of sites in the (magnetic) unit cell, and $\hat{H}(\bs{k})$
is the following generic Hamiltonian matrix in reciprocal space:
\begin{align}
  \label{eq:18}
  \hat{H}(\bs{k})&=\sum_{a,b=1}^N\sum_{\mu=0}^3 {\sf h}_{ab}^\mu({\bs k})
\hat{E}_{ab}\hat{\sigma}^\mu,
\end{align}
where $\hat{E}_{ab}$ is the $N\times N$ matrix with a $1$ at position
$ab$ and zeros everywhere else, i.e.\ with matrix elements
\begin{equation}
  \label{eq:13}
  (\hat{E}_{ab})_{ij}\equiv \delta_{ai}\delta_{bj},
\end{equation}
where we use a `hat' on $\hat{E}$ in order to emphasize
$\hat{E}_{ab}$ is a matrix rather than a matrix
{\em element}, $\hat{\sigma}^0={\rm Id}_2$ is the identity matrix, 
$\hat{\sigma}^{1,2,3}$ are the three Pauli matrices, with
\begin{align}
  \label{eq:300}
  {\sf h}_{a\neq b}^\mu(\bs{k})&=\sum_{\eta=1}^{{\rm n}_{ab}} t_{ab,(\eta)}^\mu
                                 e^{i\bs{k}\cdot\bs{e}_{ab}^{(\eta)}},\\
  {\sf h}_{aa}^{\mu\neq0}(\bs{k})&=K_a^\mu S_a^\mu+\sum_{\eta=1}^{{\rm n}_{aa}} t_{aa,(\eta)}^\mu
                                   e^{i\bs{k}\cdot\bs{e}_{aa}^{(\eta)}}\nonumber,\\
{\sf h}_{aa}^{0}(\bs{k})&=\sum_{\eta=1}^{{\rm n}_{aa}} t_{aa,(\eta)}^\mu
e^{i\bs{k}\cdot\bs{e}_{aa}^{(\eta)}}\nonumber,
\end{align}
where $ab$ is a bond between
sublattices $a$ and $b$ and the sum over $\eta$ runs over the ${\rm
  n}_{ab}$ $ab$-type bonds
with a nonzero hopping and separated by
$\bs{e}_{a\neq b}^{(\eta)}=-\bs{e}_{ba}^{(\eta)}$, $K_a^\mu$ parametrizes the
Kondo coupling on sublattice $a$, and hermiticity imposes
$({\sf h}_{ab}^\mu)^*={\sf h}_{ba}^\mu$ so that we can set
$(t_{ab,(\eta)}^\mu)^*=t_{ba,(\eta)}^\mu$ (more precisely, it is possible to make such
choices of $\eta$). For convenience, we also define ${\sf
  h}_{ab,(\eta)}^\mu(\bs{k})$ such that ${\sf
  h}_{ab}^\mu(\bs{k})=\sum_\eta{\sf h}_{ab,(\eta)}^\mu(\bs{k})$, i.e.\ 
${\sf h}_{ab,(\eta)}^\mu(\bs{k})=t_{ab,(\eta)}^\mu
e^{i\bs{k}\cdot\bs{e}_{ab}^{(\eta)}}$ for $\eta=1,..,{\rm n}_{ab}$,
and ${\sf h}_{aa,(0)}^{\mu\neq0}(\bs{k})=K_a^\mu S_a^\mu$. We also note
\begin{equation}
  \label{eq:33}
    \partial_{k_\alpha}{\sf h}_{ab}^\mu=i \sum_\eta
                                    e^{\alpha}_{ab,(\eta)}t_{ab,(\eta)}^\mu
                                    e^{i\bs{k}\cdot\bs{e}_{ab}^{(\eta)}}
\end{equation}
for future reference. 
Note that it is the decomposition Eq.~\eqref{eq:18} in sublattice
$a,b$ and spin $\mu$ indices and the separate traces which will allow
to relate the Berry curvature to real space (sublattice) and spin quantities.

As applications, we will
consider the nearest-neighbor kagom\'e lattice and the
three-sublattice nearest-neighbor triangular lattice with a triangular
basis, see Fig.~\ref{fig:three}. In
both cases, $M=2N=6$. 
In the
kagom\'e lattice
case, on each $ab=12,23,31$-type bond, $\eta=1,2=\mp$. 
In the triangular lattice case, $\eta=1,2,3$ for each $ab$-bond
type. 

{\bf {\em Formulas for the Berry curvature in terms of the Hamiltonian.---}}We now turn to the expressions of the Berry curvature
and quantum metric in terms of the Hamiltonian. As mentioned above, we
make use of band projectors in
expressing the Berry curvature, and in particular rederive formulas
present in Refs.~\cite{graf2021,grafthesis} using matrices rather than
Bloch vectors. Using the formula $G_{(n)}^{\alpha\beta}={\rm Tr}[\partial_\alpha
\hat{P}_{(n)}(1-\hat{P}_{(n)})\partial_\beta \hat{P}_{(n)}]$ for the quantum geometric tensor in band $n$ with
``directions'' $\alpha,\beta$ where $\partial_\alpha\equiv\partial_{k_\alpha}$,
and $\hat{P}_{(n)}=|u_n(\bs{k})\rangle\langle u_n(\bs{k})|$ the
projector into band $n$ (we will look only away from band degeneracies), we can write
$G_{(n)}^{\alpha\beta}=\Gamma^{\alpha\beta}_{(n)}-\frac{i}{2}\Omega^{\alpha\beta}_{(n)}$
\cite{provost1980,resta2011,graf2021}, where $\Gamma^{\alpha\beta}_{(n)}\equiv{\rm Re}G_{(n)}^{\alpha\beta}$ and
$\Omega^{\alpha\beta}_{(n)}=-2{\rm Im}G_{(n)}^{\alpha\beta}$ are the quantum metric \cite{provost1980} and
Berry curvature (note that in three-dimensions we can use equivalently
two indices $\alpha\beta$ or one perpendicular direction index $\gamma$), respectively.

Then, using $\hat{H}=\sum_n\varepsilon_n \hat{P}_{(n)}$
($\varepsilon_n$ is the energy in band $n$) and $\hat{P}_{(n)}=\prod_{m\neq n}(\hat{H}-\varepsilon_m)/\prod_{m\neq
  n}(\varepsilon_n-\varepsilon_m)$ 
one can show that
\begin{equation}
  \label{eq:101}
  \hat{P}_{(n)}=\sum_{r=0}^{M-1}\ell^{(n)}_{r}\hat{H}^{r},
\end{equation}
where $\ell^{(n)}_{r}$ are prefactors which depend only on
$\varepsilon_n$ and ${\rm Tr}\hat{H}^{r'}$, and $\hat{H}^0\equiv{\rm Id}_M$. 
The exact expressions of the prefactors $\ell_r^{(n)}$
are given in Appendix~\ref{sec:proj-poly} (Eq.~\eqref{eq:127}). What
is most 
important is that (i) the sum in Eq.~\eqref{eq:101} {\em terminates} and
(ii) $\hat{P}_{(n)}$ can be entirely expressed in terms of
$\varepsilon_n(\bs{k})$ and the hamiltonian matrix $\hat{H}(\bs{k})$, i.e.\ in particular
no eigenvectors are required and only the $n$th eigenvalue of $\hat{H}$ must be calculated. The
finiteness of the sum in $\hat{P}_{(n)}$ means we can organize the
terms in ``powers'' of $\hat{H}$, from $3$ to $3(M-1)$ 
in the case
of the Berry curvature and from $2$ to $3(M-1)$ in the case of the
quantum metric. The maximum number of matrix elements in a product therefore grows like the volume of
the unit cell.

\begin{figure}[htbp]
    \centering
    \subfigure[]{\includegraphics[width=0.25\columnwidth]{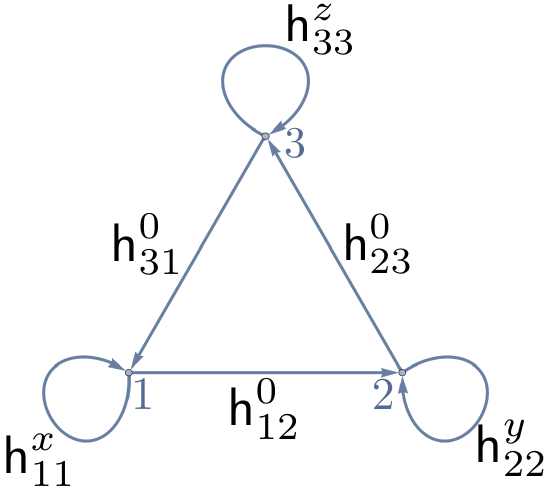}} 
    \subfigure[]{\includegraphics[width=0.5\columnwidth]{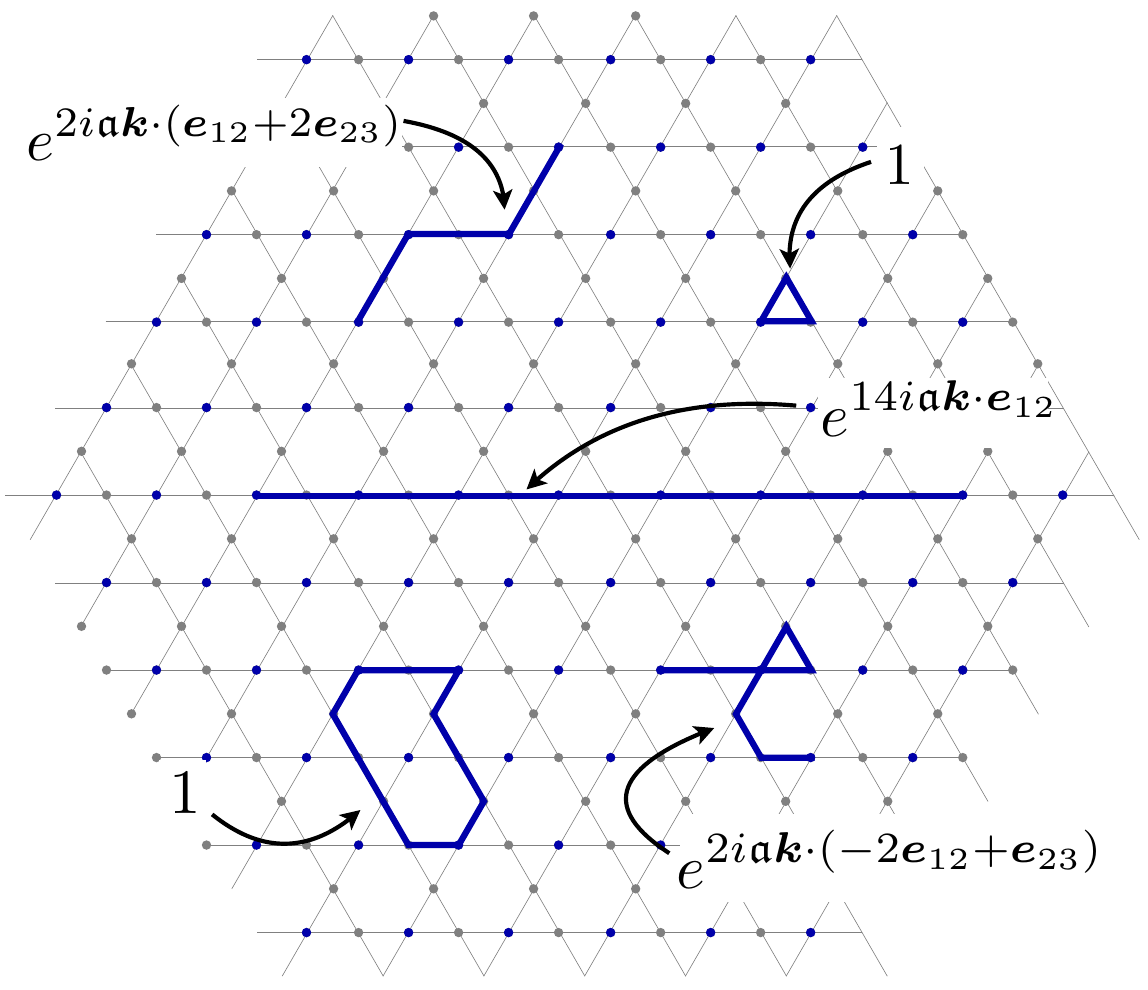}} 
    \caption{(a) Graphical representation of a length-6 loop (e.g.\
      $r_1=r_2=r_3=2$) in a three-sublattice {\em unit cell}. This
      produces, among others a $\Lambda_{\{1,3\}}$ contribution (see Eqs.~(\ref{eq:63},\ref{eq:17}). (b) Loops of unit cell indices
      become strings on the lattice, with both start and end points belonging
      to the same sublattice. Here we are not showing on-site loops 
      as
    in (a), but each site may host any number. In
    $\Omega^{\alpha\beta}_{(n)}$, each string is weighed by the factor
    shown multiplied by
    $e^\alpha_{a_1a_2(\eta_1)}e^\beta_{a_pa_{p+1}(\eta_p)}$ and any
    on-site factors. The straight line represents the contribution with the maximum
    possible extent.}
    \label{fig:loopsandstrings}
\end{figure}

Moreover, as noted in Refs.~\cite{graf2021,grafthesis}, in the case of the Berry curvature,
``orthogonality'' relations (see Appendix~\ref{sec:berry-curv-terms})
allow one to rewrite the Berry curvature as
\begin{align}
  \label{eq:103}
  \Omega_{(n)}^{\alpha\beta}&=-i
  \sum_{r_1,r_2,r_3}\ell^{(n)}_{r_1}\ell^{(n)}_{r_2}\ell^{(n)}_{r_3}{\rm Tr}\left[[\partial_\beta(\hat{H}^{r_3}),\partial_\alpha(\hat{H}^{r_1})] \hat{H}^{r_2}\right]
\end{align}
(the sums over the $r_i$ run from $1$ to $2N-1$), i.e.\ when applying the chain rule on
$\partial_{\alpha/\beta}\hat{P}_{(n)}$ using Eq.~\eqref{eq:101} in $\Omega^{\alpha\beta}_{(n)}$, only the terms with the derivatives acting on the
$\hat{H}^{r}$ survive, and not those with derivatives acting on
the coefficients $\ell^{(n)}_{r}$. Furthermore, using the chain rule on
Eq.~\eqref{eq:103}, we can write, using
$\sum_{p_1=0}^{r_1-1}\sum_{p_3=0}^{r_3-1}\cdots\sim\sum_{p=0}^{r_1+r_3-2}\xi_{r_1,r_3}(p)\cdots$
with 
$\xi_{\{r_i\}}(p)={\rm min}\left[{\rm
  min}(r_1,r_3),(r_1+r_3)/2-|p-(R+r_2+2)/2|\right]\in\mathbb{N}$ when $\cdots$ depends on $p_1+p_3$ only,
\begin{align}
  \label{eq:63}
  \Omega^{\alpha\beta}_{(n)}&=2\sum_{r_{1,2,3}}\ell_{r_1}^{(n)}\ell_{r_2}^{(n)}\ell_{r_3}^{(n)}\\
  &\quad\qquad\sum_{p=r_2+2}^{R}\xi_{\{r_i\}}(p){\rm
  Im}{\rm
  Tr}[\partial_\alpha\hat{H}\hat{H}^{p-2}\partial_\beta\hat{H}\hat{H}^{R-p}]\nonumber,
\end{align}
where we defined $R\equiv r_1+r_2+r_3$, and in the sum we have
$q_1\equiv p-2 \in\llbracket r_2-1, r_1+r_2-2\rrbracket$ and
  $q_2\equiv R-p\in\llbracket r_3, r_1+r_3-1\rrbracket$. Because of the antisymmetry of
$\Omega^{\alpha\beta}_{(n)}$ under $\alpha\leftrightarrow\beta$, many
terms in the sums in Eq.~\eqref{eq:63} cancel. This is in particular
true because $\Lambda_{(q_1,q_2)}^{\alpha\beta}\equiv {\rm
  Tr}[\partial_\alpha\hat{H}\hat{H}^{q_1}\partial_\beta\hat{H}\hat{H}^{q_2}]=-\Lambda_{(q_2,q_1)}^{\alpha\beta}$,
so that in $\Omega^{\alpha\beta}_{(n)}$, if
$\Lambda_{(q_1,q_2)}^{\alpha\beta}$ and
$\Lambda_{(q_2,q_1)}^{\alpha\beta}$ both appear with the same $\ell$
prefactors, they cancel against one another. In Eq.~\eqref{eq:40} we
provide the resulting explicit
expression for the
Berry curvature of a three-sublattice system.

{\bf {\em Formulas for the Berry curvature in terms of Hamiltonian matrix elements.---}}Let us now write powers of $\hat{H}$ using the matrix elements defined
above. We have 
$\hat{E}_{ab}\hat{E}_{cd}=\delta_{bc}\hat{E}_{ad}$,
$\hat{\sigma}^\mu\hat{\sigma}^\nu=\sum_{\rho=0}^3g_{\mu\nu\rho}\hat{\sigma}^\rho$, with
$g_{\mu\nu\rho}=d_{\mu\nu\rho}+if_{\mu\nu\rho}$ where, for $\mu,\nu,\rho=0,..,3$,
\begin{align}
  \label{eq:403}
  d_{\mu\nu\rho}&=\delta_{(\mu\nu}\delta_{\rho)0}-2\delta_{\mu0}\delta_{\nu0}\delta_{\rho0}\nonumber,\\
  f_{\mu\nu\rho}&=\epsilon_{\mu\nu\rho}. 
\end{align}
Here
$\delta_{(\mu\nu}\delta_{\rho)0}\equiv\delta_{\mu\nu}\delta_{\rho0}+\delta_{\rho\mu}\delta_{\nu}+\delta_{\nu\rho}\delta_{\mu0}$
is the symmetrized sum and $\epsilon_{\mu\nu\rho}$ is the
three-dimensional Levi-Civita tensor where implicitly
$\epsilon_{\mu\nu\rho}=0$ if $\mu$, $\nu$ or $\rho$ is zero. In turn,
\begin{equation}
  \label{eq:501}
  \hat{H}^r=\sum_{\{\mu_i\},\{\rho_j\}}g_{\mu_1\mu_2\rho_2}\cdots g_{\rho_{r-1}\mu_r\rho_r}\hat{\sf
    H}_{\mu_1}\cdots\hat{\sf H}_{\mu_r}\hat{\sigma}_{\rho_r},
\end{equation}
where $\hat{\sf H}_{\mu}$ is the $N\times N$ matrix with matrix
elements $(\hat{\sf H}_{\mu})_{ab}={\sf h}_{ab}^\mu$, 
and we have $(\hat{\sf H}_\mu)^\dagger=\hat{\sf
  H}_\mu$. Finally (recall $R=r_1+r_2+r_3$), 
\begin{widetext}
\begin{align}
  \label{eq:37}
  \Omega_{(n)}^{\alpha\beta}=2\sum_{r_1,r_2,r_3=1}^{2N-1}\mathcal{L}_{r_1,r_2,r_3}^{(n)}
  \sum_{\{\mu_i\}_{i=1,..,R}}\!\!\!{\rm Im}\left(\mathcal{G}_{\mu_1\cdots\mu_R}\sum_{p=r_2+2}^{R}\xi_{\{r_i\}}(p)\mathcal{H}_{\mu_1\cdots\mu_R}^{[p]}\right).
\end{align}
Here, 
$\mathcal{L}^{(n)}_{r_1,r_2,r_3}=\ell_{r_1}^{(n)}\ell_{r_2}^{(n)}\ell_{r_3}^{(n)}$
is the product of projector prefactors defined in Eq.~\eqref{eq:101},
$\mathcal{G}_{\mu_1\cdots\mu_R}={\rm Tr}[\hat{\sigma}_{\mu_1}\cdots\hat{\sigma}_{\mu_R}]=\frac{1}{2}\sum_{\{\rho_i\}_{i=1,..,R}}g_{\rho_{R}\mu_1\rho_1}\cdots
g_{\rho_{R-1}\mu_{R}\rho_{R}}$ (see
Appendix~\ref{sec:general-relations}) is the contraction of Lie algebra
structure constants defined in
Eq.~\eqref{eq:403} and can be tabulated
once and for all. It is $\mathcal{G}$ which will entirely fix
the ``geometric'' structure of the terms in the Berry curvature, i.e.\
determine which contributions such as
$\vec{S}_i\cdot(\vec{S}_j\times\vec{S}_k)$, $\vec{S}_i\cdot\vec{S}_j$,
$\vec{t}_{ij}\cdot(\vec{t}_{kl}\times\vec{t}_{mq})$,
$\vec{t}_{ij}\cdot(\vec{S}_{k}\times\vec{S}_{l})$ etc.\ appear in
$\Omega$ (see below). Finally 
\begin{align}
  \label{eq:19}
 \mathcal{H}_{\mu_1\cdots\mu_R}^{\alpha\beta,[p]}&= {\rm Tr}[\partial_\alpha\hat{\sf
    H}_{\mu_1}\hat{\sf H}_{\mu_2}\cdots\partial_\beta\hat{\sf
  H}_{\mu_{p}}\hat{\sf H}_{\mu_{p+1}}\cdots\hat{\sf H}_{\mu_{R}}],
\end{align}
where $p$ labels the index where $\partial_\beta$
acts. For the
kagom\'e and three-sublattice triangular lattices, $1\leq r_i\leq 5$,
and so $3\leq R\leq15$.

It is interesting to make the structure of the Berry curvature as a sum of
``loops'' (resp.\ strings with ends on identical sublattices) of varying lengths
within the unit cell---Fig.~\ref{fig:loopsandstrings}(a)---(resp.\ on the lattice---Fig.~\ref{fig:loopsandstrings}(b)) more obvious. Using Eq.~\eqref{eq:300}, we
write 
\begin{align}
  \label{eq:23}
\mathcal{H}_{\mu_1\cdots\mu_R}^{\alpha\beta,[p]}
  &=\!\!\!\!\sum_{\{a_1,a_2,a_p,a_{p+1}\}}\!\!\!\!J_{a_1,a_2,a_p,a_{p+1}}^{\alpha\beta|\mu_1\mu_p}(\hat{\sf
    H}_{\mu_2}\cdots\hat{\sf H}_{\mu_{p-1}})_{a_2a_{p}}(\hat{\sf
    H}_{\mu_{p+1}}\cdots\hat{\sf H}_{\mu_{R}})_{a_{p+1}a_1},
\end{align}
and we note $(\hat{\sf
    H}_{\mu_2}\cdots\hat{\sf H}_{\mu_{p-1}})_{a_2a_{p}}={\sf
    h}^{\mu_2}_{a_2a_3}\cdots{\sf h}^{\mu_{p-1}}_{a_{p-1}a_{p}}$, and
\begin{align}
  \label{eq:503}
J_{a_1,a_2,a_p,a_{p+1}}^{\alpha\beta|\mu_1\mu_p} &=-\sum_{\eta_1,\eta_p}e_{a_1a_2(\eta_1)}^\alpha e_{a_pa_{p+1}(\eta_p)}^\beta{\sf
    h}^{\mu_1}_{a_1a_2,(\eta_1)}{\sf
  h}^{\mu_{p}}_{a_{p}a_{p+1},(\eta_p)} 
\end{align}
are the products of the matrix elements for the terms on which
$\partial_{\alpha,\beta}$ act.  We may therefore write
\begin{align}
  \label{eq:41}
  \mathcal{H}_{\mu_1\cdots\mu_R}^{\alpha\beta,[p]}
  &=-\sum_{\{a_i\}}\sum_{\eta_1,\eta_p}[e_{a_1a_2(\eta_1)}^\alpha
    e_{a_pa_{p+1}(\eta_p)}^\beta]\;{\sf
    h}^{\mu_1}_{a_1a_2,(\eta_1)}{\sf h}^{\mu_2}_{a_2a_3}\cdots{\sf
    h}^{\mu_p}_{a_pa_{p+1},(\eta_p)}{\sf
    h}^{\mu_{p+1}}_{a_{p+1}a_{p+2}}\cdots{\sf
    h}^{\mu_{R}}_{a_Ra_{1}}\nonumber\\
  &=-\sum_{\{a_i\}}\sum_{\{\eta_i\}}[e_{a_1a_2(\eta_1)}^\alpha
    e_{a_pa_{p+1}(\eta_p)}^\beta]\;t^{\mu_1}_{a_1a_2(\eta_1)}t^{\mu_2}_{a_2a_3(\eta_2)}\cdots
    t^{\mu_{R}}_{a_Ra_{1}(\eta_R)}e^{i\bs{k}\cdot({\bs e}_{a_1a_2(\eta_1)}+{\bs
    e}_{a_2a_3(\eta_2)}+\cdots+
    {\bs e}_{a_Ra_{1}(\eta_R)})}.
\end{align}
\end{widetext}
We also note that the second line of Eq.~\eqref{eq:41} involves sums of $e^{i\bs{k}\cdot\bs{\mathcal{S}}}$, where
\begin{equation}
  \label{eq:42}
  \bs{\mathcal{S}}=\bs{e}_{a_1a_2(\eta_1)}+\cdots+\bs{e}_{a_Ra_1(\eta_R)}.
\end{equation}
Because the `strings' $\bs{\mathcal{S}}$ always ``start'' and ``end'' on a given
sublattice, $\bs{\mathcal{S}}$ is always a {\em Bravais} lattice
vector, 
$\bs{\mathcal{S}}=\sum_{i=1}^d\sum_{n_i} n_i\bs{A}_i$,
$n_i\in\mathbb{Z}$, where the $\bs{A}_i$ are elementary Bravais lattice vectors. For the
kagom\'e lattice where the distance between two nearest-neighbor sites
is $\mathfrak{a}$, we have for example $\bs{A}_1=2\mathfrak{a}\bs{e}_{12}$ and
$\bs{A}_2=2\mathfrak{a}\bs{e}_{13}$. In turn, while the $e^{i\bs{k}\cdot\bs{\mathcal{S}}}$ terms take different
prefactors, as $R$ becomes larger, the sums of these terms become
increasingly peaked around the $\bs{k}$ values where
$\bs{k}\cdot\bs{\mathcal{S}}=0\;[2\pi]$, i.e.\ at the reciprocal
Bravais 
lattice vectors, $\bs{k}_{\rm peak}=\sum_{i=1}^d\sum_{x_i}
x_i\bs{B}_i$, $x_i\in\mathbb{Z}$, with e.g.\
$\bs{B}_1=\frac{\pi}{\mathfrak{a}}\frac{\bs{e}_{13}\times
  \bs{\hat{z}}}{\bs{e}_{12}\cdot(\bs{e}_{13}\times \bs{\hat{z}})}$, $\bs{B}_2=\frac{\pi}{\mathfrak{a}}\frac{\bs{\hat{z}}\times\bs{e}_{12}}{\bs{e}_{12}\cdot(\bs{e}_{13}\times \bs{\hat{z}})}$ for the kagom\'e lattice.

\begin{figure}[htbp]
  \centering
  \includegraphics[width=0.5\columnwidth]{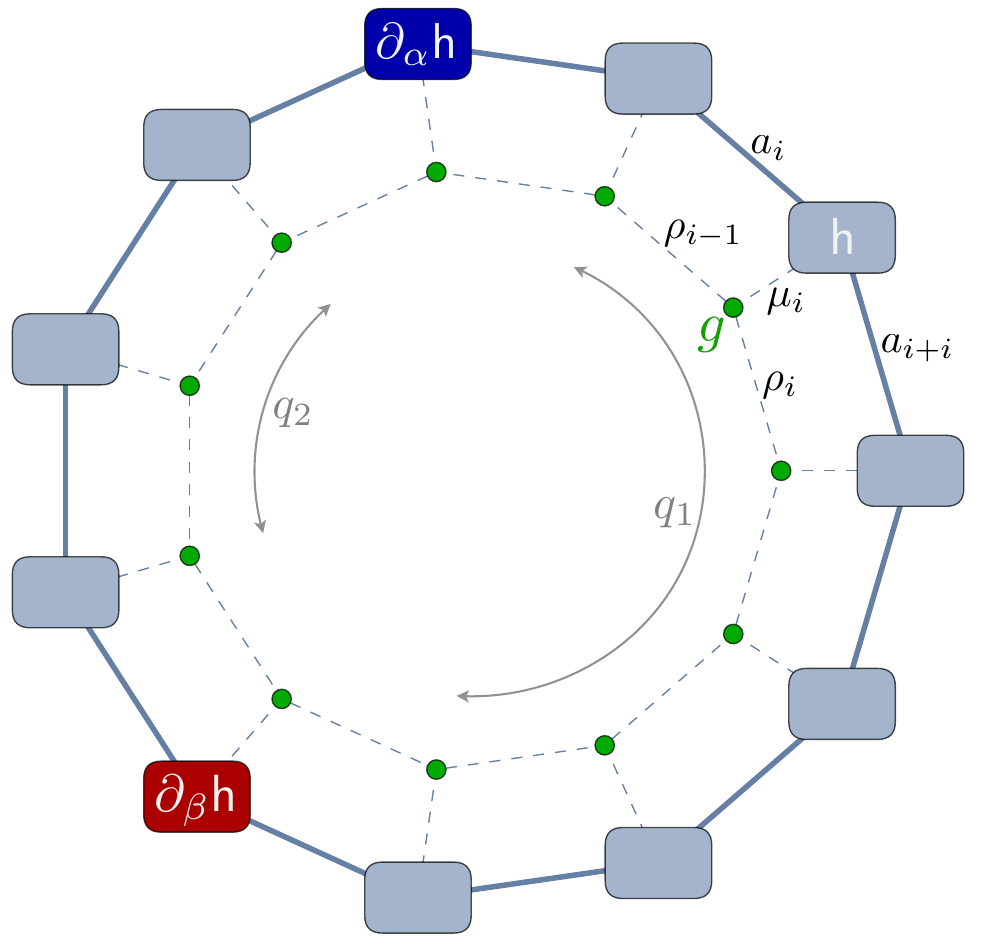}
  \caption{Graphical `tensor network' representation of one of the terms in
    $\Omega_{(n)}^{\alpha\beta}$, Eq.~\eqref{eq:103}, with
    $(q_1,q_2)=(6,3)$ present for example in the sum for $r_1=3$,
    $r_2=4$ and $r_3=4$. The light blue rectangles represent $\hat{\sf
      H}$ while the dark blue and dark red rectangles
    $\partial_{\alpha/\beta}\hat{\sf H}$, respectively, while the green circles represent the
    `structure constant' $g=d+if$. Lines represent contractions of 
    $\mu$ (between a rectangle and a circle) and $\rho$ (between
    circles) spin indices and $a$ site indices (between
    rectangles). }
  \label{fig:contraction}
\end{figure}

Finally, we note that $\Omega_{(n)}^{\alpha\beta}$ can be represented as a sum of
contractions of tensors, which we show graphically in
Fig.~\ref{fig:contraction}.


{\bf {\em Nontrivial case where the Berry curvature vanishes for any spin
  configuration.---}}If (for $a\neq b$) ${\sf
  h}_{ab,(\eta)}^\mu=t_0 e^{i\bs{k}\cdot\bs{e}_{ab(\eta)}}$
(i.e.\ $t_{ab(\eta)}^\mu$ is independent of $ab(\eta)$ and of $\mu$), then it is useful to define $\bs{I}_{ab}\equiv\sum_\eta
\bs{e}_{ab(\eta)}e^{i\bs{k}\cdot\bs{e}_{ab(\eta)}}=-\bs{I}_{ba}^*$ (no
summation over $a,b$!). In the case of
the triangular lattice
\begin{align}
  \label{eq:505}
  \bs{I}_{ab}^{\rm t}&=\upsilon_{ab}(\bs{e}_{12(0)}e^{i
    \upsilon_{ab}\bs{k}\cdot\bs{e}_{12(0)}}\\
  &\quad\qquad+\bs{e}_{23(0)}e^{i \upsilon_{ab}\bs{k}\cdot\bs{e}_{23(0)}}+\bs{e}_{31(0)}e^{i \upsilon_{ab}\bs{k}\cdot\bs{e}_{31(0)}}),\nonumber
\end{align}
where $\upsilon_{12}=\upsilon_{23}=\upsilon_{31}$, $\upsilon_{ba}=-\upsilon_{ab}=\pm1$. This entails that ${\bs I}_{12}^{\rm t}={\bs I}_{23}^{\rm t}={\bs
  I}_{31}^{\rm t}=-({\bs I}_{21}^{\rm t})^*=-({\bs I}_{32}^{\rm
  t})^*=-({\bs I}_{13}^{\rm t})^*$. Using the latter, the hermiticity
of $\hat{\sf H}_\mu$, and 
$\Omega^{\alpha\beta}=-\Omega^{\beta\alpha}$ and
$\Omega^{\alpha\beta}\in\mathbb{R}$, one can show
that the Berry curvature vanishes for any configuration of the
spins. This is an important result of this work. We provide details of the derivation in
Appendix~\ref{sec:triangular}. Note that this does {\em not} apply for
example in the case of the kagom\'e lattice where $\bs{I}_{ab}^{\rm
  k}=\upsilon_{ab}\bs{e}_{ab}(e^{i\bs{k}\cdot\bs{e}_{ab}}-e^{-i\bs{k}\cdot\bs{e}_{ab}})^{\upsilon_{ab}}$
and so e.g.\ $\bs{I}_{12}^{\rm
  k}\neq \bs{I}_{23}^{\rm
  k}$.

{\bf {\em Berry curvature as a ``polynomial'' of geometric elements of the
  spin (and/or hopping vector) texture in the unit cell.---}}We now
investigate the structure of the contractions between $g$ tensors and
$\hat{\sf H}_\mu$ matrices in order to explicitly express the Berry
curvature as a ``polynomial'' in terms such as
$\vec{S}_i\cdot\vec{S}_j$, $\vec{S}_i\cdot(\vec{S}_j\times\vec{S}_k)$, 
and their products and powers, with $\bs{k}$-dependent coefficients
that can be exactly and explicitly computed. [Note that we use
quotation marks around ``polynomial'' because the $\ell_r^{(n)}$
coefficients also in principle depend on the spins through
$\varepsilon_n$ and ${\rm Tr}\hat{H}^{r'}$.] In other words, for
three-sublattice systems without spin-orbit coupling, we can write:
\begin{widetext}
\begin{equation}
  \label{eq:3}
  \Omega^{\alpha\beta}_{(n)}(\mathbf{k})=\sum_{r_1,r_2,r_3}\ell^{(n)}_{r_1}\ell^{(n)}_{r_2}\ell^{(n)}_{r_3}\mathcal{P}^{\alpha\beta}_{i_{12},i_{23},i_{31},i_{123}} (\vec{S}_1\cdot\vec{S}_2)^{i_{12}}(\vec{S}_2\cdot\vec{S}_3)^{i_{23}}(\vec{S}_3\cdot\vec{S}_1)^{i_{31}}(\chi_{123})^{i_{123}},
\end{equation}
with $\chi_{123}=\vec{S}_1\cdot(\vec{S}_2\times\vec{S}_3)$, $i_{12,23,31,123}\in\mathbb{N}$ and
$i_{\rm tot}=2(i_{12}+i_{23}+i_{31})+3i_{123}\leq13$, and the $\mathcal{P}$'s are
functions of only $\bs{k}$ (and $t_0,K$) that can be determined
analytically. Importantly, in a spin-orbit coupling-free system, all
contributions to the Berry curvature involve between $3$ and
$3(2N-1)-2$ powers of $K$. Longer strings are only associated with higher $i_{\rm tot}$ and
in turn higher powers of the Kondo coupling, so that in the
weak-Kondo-coupling regime contributions from shorter strings will
dominate the Berry curvature.

In the presence of
spin-orbit coupling, one must include in the sum polynomials of all
the following terms
\begin{align}
  \label{eq:15}
  &(\vec{S}_1\cdot\vec{S}_2), (\vec{S}_2\cdot\vec{S}_3), (\vec{S}_3\cdot\vec{S}_1)\nonumber\\
 & (\chi_{123}), \nonumber\\
 & (\vec{t}_{12}\cdot\vec{t}_{23}), (\vec{t}_{23}\cdot\vec{t}_{31}), (\vec{t}_{31}\cdot\vec{t}_{12}) \nonumber\\
 & (\vec{t}_{12}\cdot(\vec{t}_{23}\times\vec{t}_{31})), \nonumber\\
 & (\vec{S}_1\cdot\vec{t}_{12}), (\vec{S}_1\cdot\vec{t}_{23}),
  (\vec{S}_1\cdot\vec{t}_{31}), (\vec{S}_2\cdot\vec{t}_{12}),
  (\vec{S}_2\cdot\vec{t}_{23}), (\vec{S}_2\cdot\vec{t}_{31}),
  (\vec{S}_3\cdot\vec{t}_{12}), (\vec{S}_3\cdot\vec{t}_{23}),
  (\vec{S}_3\cdot\vec{t}_{31}), \nonumber\\
&  (\vec{t}_{12}\cdot(\vec{S}_1\times\vec{S}_2)),
  (\vec{t}_{23}\cdot(\vec{S}_1\times\vec{S}_2)), (\vec{t}_{31}\cdot(\vec{S}_1\times\vec{S}_2)), (\vec{t}_{12}\cdot(\vec{S}_2\times\vec{S}_3)),
  (\vec{t}_{23}\cdot(\vec{S}_2\times\vec{S}_3)),
                                             (\vec{t}_{31}\cdot(\vec{S}_2\times\vec{S}_3)),
                                             (\vec{t}_{12}\cdot(\vec{S}_3\times\vec{S}_1)),\nonumber\\
  &\qquad\qquad
  (\vec{t}_{23}\cdot(\vec{S}_3\times\vec{S}_1)), (\vec{t}_{31}\cdot(\vec{S}_3\times\vec{S}_1)), \nonumber\\
&  (\vec{S}_1\cdot(\vec{t}_{12}\times\vec{t}_{23})),\cdots,
\end{align}
\end{widetext}
where $\vec{t}_{ab}\equiv(t^x_{ab},t^y_{ab},t^z_{ab})$ and we
suppressed the $\eta$ subscripts for clarity (other ``similar''
combinations are also possible involving $(t_{ab}^\mu)^*$ if the $t$'s
are complex). 

\begin{figure}[htbp]
  \centering
  \includegraphics[width=.7\columnwidth]{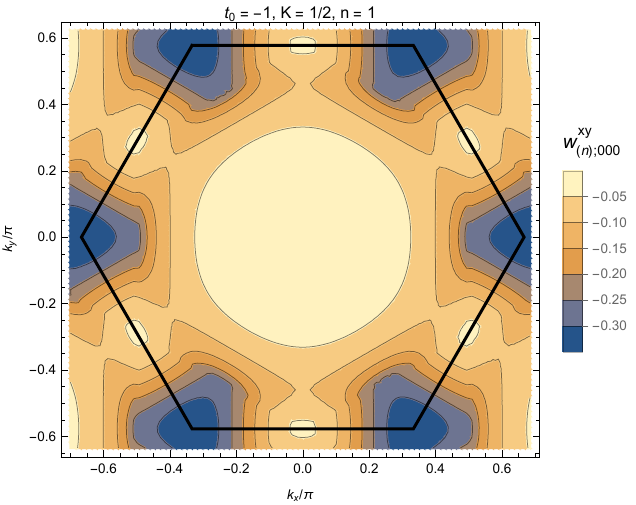}
  \caption{$w_{(1);000}^{xy}(\bs{k})$ as defined in Eq.~\eqref{eq:24},
    $\Omega_{(1)}^{xy}(\bs{k})=\chi_{123}w_{(1);000}^{xy}(\bs{k})$ with $\chi_{123}=\vec{S}_1\cdot(\vec{S}_2\times\vec{S}_3)$, for
    the spin-orbit-coupling-free model on the kagom\'e lattice with
    $t_0=-1,K=1/2$ with $\vec{S}_1\perp\vec{S}_2\perp\vec{S}_3\perp\vec{S}_1$. }
  \label{fig:finalplot}
\end{figure}

For (relative) simplicity, we focus on a nearest-neighbor-only case without spin-orbit
coupling, $t^0_{a\neq b}=t_0$, $\vec{t}_{a\neq b}=\vec{0}$, $K_a^\mu=K$, and such that no nearest-neighbors belong to the same
sublattice (this ensures that the terms on the diagonal blocks of the
Hamiltonian matrix are Kondo-coupled spins only). Since $g_{\mu\nu0}=\delta_{\mu\nu}$, in that case, the
zero-components, $\mu=0$, correspond to intersublattice hopping
terms (block off-diagonal, $a\neq b$), while the
$\mu=1,2,3$ components are the Kondo-coupled spin terms (block diagonal, $a=b$). 
$\mathcal{H}$ comes down to summing the exponentials of all the paths one can take
in a fixed (and {\em bounded}) number of (here, nearest-neighbor) steps from one
sublattice point to another, see Fig.~\ref{fig:loopsandstrings}(b),
and we find that $\Omega^{\alpha\beta}$ (away from band crossings)
simply equals
\begin{align}
  \label{eq:24}
  &\Omega^{\alpha\beta}_{(n)}(\bs{k})=\chi_{123}\\
  &\sum_{0\leq i_{12}+i_{23}+i_{31}\leq2}\! \! \! \! \! \! \! \! \! \! \! \! \! \! \! w^{\alpha\beta}_{(n);i_{12},i_{23},i_{31}}(\bs{k}) (\vec{S}_1\cdot\vec{S}_2)^{i_{12}}(\vec{S}_2\cdot\vec{S}_3)^{i_{23}}(\vec{S}_3\cdot\vec{S}_1)^{i_{31}},\nonumber
\end{align}
where the $w$'s are functions of $\bs{k}$, which we obtain
analytically (we list the first few $\Lambda_{(q_1,q_2)}^{xy}$'s in
Appendix~\ref{sec:three-subl-struct}, Eq.~\eqref{eq:17}).
In Figure~\ref{fig:finalplot} we plot $w^{xy}_{(n);000}$ for the kagom\'e lattice for
$t_0=-1,K=1/2$ and $n=1$ (and $\vec{S}_1=(1,0,0)$, $\vec{S}_2=(0,1,0)$
and $\vec{S}_3=(0,0,1)$, whose specification is {\em only} required to compute
$\epsilon_n(\bs{k})$ because it appears in the $\ell^{(n)}$'s).

{\bf {\em Discussion.---}}In this manuscript, we have provided an {\em exact} method to
compute the Berry curvature analytically as a {\em finite} sum over
{\em finite} strings on the lattice (this is in contrast, for example,
with the claims in e.g.\ Ref.~\cite{tatara2002}), in particular in electronic
systems Kondo-coupled to spins. We made no assumptions on the strengh of
the Kondo coupling, `double exchange' \cite{karplus1954}, long-wavelength limits,
or on the size of the magnetic unit cell. We expect that this will
allow a more accurate interpretation of experiments as well as easier
and more accurate calculations of the Berry curvature and associated
Hall conductivity than those
accessible now \cite{fukui2005,weisse2006}. Of course, beyond Berry
curvature effects, skew-scattering \cite{ishizuka2021,nagaosa2010} may
also contribute to $\sigma^{xy}_{\rm H}$ in real systems, but our
exact derivation of the intrinsic effects should allow to better
distinguish the contributions.

We explicitly applied our formalism to three-sublattice systems
without spin-orbit coupling and showed that the Berry curvature
vanished for the triangular lattice, regardless of the orientation of
the spins. On the kagom\'e lattice (and any other
three-sublattice system), we found that all contributions to the
Berry curvature included a single power of the scalar chirality
of the spins $\chi_{123}=\vec{S}_1\cdot(\vec{S}_2\times\vec{S}_3)$ on
{\em one} sublattice, and some of the ones involving higher
powers of the Kondo coupling also included $\vec{S}_i\cdot\vec{S}_j$ elements. 




This formalism allows for many extensions as it is merely an analytical way to
relate energies and combinations of matrix elements to band
quantities. In particular, higher-spin systems, systems with fluctuating
spins, systems coupled to degrees of freedom other than spins, the
computation of other observables \cite{graf2021}, the quantum metric, and
systems with any number of sublattices \cite{diop2022,martin2008} can be studied the same
way. In the latter case, we expect that quantities beyond
three-sublattice chiralities and/or multiple powers of the
spin-chiralities will appear because the imaginary part of
traces of Pauli matrices can then involve, for example, additional
Levi-Civita tensors. Taking the limit of the unit
cell as the entire system (in a numerical experiment for example),
where there will
be both more allowed terms (``larger loops''), but also larger sums
over small terms (``small loops''), we
believe it might be possible to use the procedure presented in this
manuscript as an expansion, perhaps in loop size, and truncate it.

As an even more immediate application, it will be interesting to explicitly explore the evolution of the
Berry curvature as a function of spin-orbit coupling, the strength
of the Kondo coupling and the size of the unit cell.

\acknowledgements This project was funded by the European Research
Council (ERC) under the European Union's Horizon 2020 research and
innovation program (Grant Agreement No.~853116, acronym
TRANSPORT). This research was also supported in part by grant NSF
PHY-2309135 to the Kavli Institute for Theoretical Physics (KITP). The
author acknowledges enlightening discussions and collaborations with
Seydou-Samba Diop during the prehistory of this work. She also thanks
Miles Stoudenmire, Olivier Gauth\'e, Cristian Batista and Leon Balents for
discussions in the final stages of this work.

\bibliography{ahe.bib}

\begin{thebibliography}{26}%
\makeatletter
\providecommand \@ifxundefined [1]{%
 \@ifx{#1\undefined}
}%
\providecommand \@ifnum [1]{%
 \ifnum #1\expandafter \@firstoftwo
 \else \expandafter \@secondoftwo
 \fi
}%
\providecommand \@ifx [1]{%
 \ifx #1\expandafter \@firstoftwo
 \else \expandafter \@secondoftwo
 \fi
}%
\providecommand \natexlab [1]{#1}%
\providecommand \enquote  [1]{``#1''}%
\providecommand \bibnamefont  [1]{#1}%
\providecommand \bibfnamefont [1]{#1}%
\providecommand \citenamefont [1]{#1}%
\providecommand \href@noop [0]{\@secondoftwo}%
\providecommand \href [0]{\begingroup \@sanitize@url \@href}%
\providecommand \@href[1]{\@@startlink{#1}\@@href}%
\providecommand \@@href[1]{\endgroup#1\@@endlink}%
\providecommand \@sanitize@url [0]{\catcode `\\12\catcode `\$12\catcode
  `\&12\catcode `\#12\catcode `\^12\catcode `\_12\catcode `\%12\relax}%
\providecommand \@@startlink[1]{}%
\providecommand \@@endlink[0]{}%
\providecommand \url  [0]{\begingroup\@sanitize@url \@url }%
\providecommand \@url [1]{\endgroup\@href {#1}{\urlprefix }}%
\providecommand \urlprefix  [0]{URL }%
\providecommand \Eprint [0]{\href }%
\providecommand \doibase [0]{http://dx.doi.org/}%
\providecommand \selectlanguage [0]{\@gobble}%
\providecommand \bibinfo  [0]{\@secondoftwo}%
\providecommand \bibfield  [0]{\@secondoftwo}%
\providecommand \translation [1]{[#1]}%
\providecommand \BibitemOpen [0]{}%
\providecommand \bibitemStop [0]{}%
\providecommand \bibitemNoStop [0]{.\EOS\space}%
\providecommand \EOS [0]{\spacefactor3000\relax}%
\providecommand \BibitemShut  [1]{\csname bibitem#1\endcsname}%
\let\auto@bib@innerbib\@empty
\bibitem [{\citenamefont {Nagaosa}(2006)}]{nagaosa2006}%
  \BibitemOpen
  \bibfield  {author} {\bibinfo {author} {\bibfnamefont {N}~\bibnamefont
  {Nagaosa}},\ }\bibfield  {title} {\enquote {\bibinfo {title} {Anomalous
  {Hall} effect: A new perspective},}\ }\href {\doibase 10.1143/JPSJ.75.042001}
  {\bibfield  {journal} {\bibinfo  {journal} {Journal of the Physical Society
  of Japan}\ }\textbf {\bibinfo {volume} {75}},\ \bibinfo {pages} {042001}
  (\bibinfo {year} {2006})},\ \Eprint
  {http://arxiv.org/abs/https://doi.org/10.1143/JPSJ.75.042001}
  {https://doi.org/10.1143/JPSJ.75.042001} \BibitemShut {NoStop}%
\bibitem [{\citenamefont {Culcer}(2024)}]{culcer2024}%
  \BibitemOpen
  \bibfield  {author} {\bibinfo {author} {\bibfnamefont {D}~\bibnamefont
  {Culcer}},\ }\bibfield  {title} {\enquote {\bibinfo {title} {The anomalous
  {Hall} effect},}\ }in\ \href {\doibase
  https://doi.org/10.1016/B978-0-323-90800-9.00006-8} {\emph {\bibinfo
  {booktitle} {Encyclopedia of Condensed Matter Physics (Second Edition)}}},\
  \bibinfo {editor} {edited by\ \bibinfo {editor} {\bibfnamefont
  {T}~\bibnamefont {Chakraborty}}}\ (\bibinfo  {publisher} {Academic Press},\
  \bibinfo {address} {Oxford},\ \bibinfo {year} {2024})\ \bibinfo {edition}
  {second edition}\ ed.,\ pp.\ \bibinfo {pages} {587--601}\BibitemShut
  {NoStop}%
\bibitem [{\citenamefont {Burkov}\ and\ \citenamefont
  {Balents}(2011)}]{burkov2011}%
  \BibitemOpen
  \bibfield  {author} {\bibinfo {author} {\bibfnamefont {A.~A.}\ \bibnamefont
  {Burkov}}\ and\ \bibinfo {author} {\bibfnamefont {Leon}\ \bibnamefont
  {Balents}},\ }\bibfield  {title} {\enquote {\bibinfo {title} {Weyl semimetal
  in a topological insulator multilayer},}\ }\href {\doibase
  10.1103/PhysRevLett.107.127205} {\bibfield  {journal} {\bibinfo  {journal}
  {Phys. Rev. Lett.}\ }\textbf {\bibinfo {volume} {107}},\ \bibinfo {pages}
  {127205} (\bibinfo {year} {2011})}\BibitemShut {NoStop}%
\bibitem [{\citenamefont {Karplus}\ and\ \citenamefont
  {Luttinger}(1954)}]{karplus1954}%
  \BibitemOpen
  \bibfield  {author} {\bibinfo {author} {\bibfnamefont {R}~\bibnamefont
  {Karplus}}\ and\ \bibinfo {author} {\bibfnamefont {J.~M.}\ \bibnamefont
  {Luttinger}},\ }\bibfield  {title} {\enquote {\bibinfo {title} {Hall effect
  in ferromagnetics},}\ }\href {\doibase 10.1103/PhysRev.95.1154} {\bibfield
  {journal} {\bibinfo  {journal} {Phys. Rev.}\ }\textbf {\bibinfo {volume}
  {95}},\ \bibinfo {pages} {1154--1160} (\bibinfo {year} {1954})}\BibitemShut
  {NoStop}%
\bibitem [{\citenamefont {Adams}\ and\ \citenamefont
  {Blount}(1959)}]{adams1959}%
  \BibitemOpen
  \bibfield  {author} {\bibinfo {author} {\bibfnamefont {E.N.}\ \bibnamefont
  {Adams}}\ and\ \bibinfo {author} {\bibfnamefont {E.I.}\ \bibnamefont
  {Blount}},\ }\bibfield  {title} {\enquote {\bibinfo {title} {Energy bands in
  the presence of an external force field: Anomalous velocities},}\ }\href
  {\doibase https://doi.org/10.1016/0022-3697(59)90004-6} {\bibfield  {journal}
  {\bibinfo  {journal} {Journal of Physics and Chemistry of Solids}\ }\textbf
  {\bibinfo {volume} {10}},\ \bibinfo {pages} {286--303} (\bibinfo {year}
  {1959})}\BibitemShut {NoStop}%
\bibitem [{\citenamefont {Thouless}\ \emph {et~al.}(1982)\citenamefont
  {Thouless}, \citenamefont {Kohmoto}, \citenamefont {Nightingale},\ and\
  \citenamefont {den Nijs}}]{thouless1982}%
  \BibitemOpen
  \bibfield  {author} {\bibinfo {author} {\bibfnamefont {D.~J.}\ \bibnamefont
  {Thouless}}, \bibinfo {author} {\bibfnamefont {M.}~\bibnamefont {Kohmoto}},
  \bibinfo {author} {\bibfnamefont {M.~P.}\ \bibnamefont {Nightingale}}, \ and\
  \bibinfo {author} {\bibfnamefont {M.}~\bibnamefont {den Nijs}},\ }\bibfield
  {title} {\enquote {\bibinfo {title} {Quantized {Hall} conductance in a
  two-dimensional periodic potential},}\ }\href {\doibase
  10.1103/PhysRevLett.49.405} {\bibfield  {journal} {\bibinfo  {journal} {Phys.
  Rev. Lett.}\ }\textbf {\bibinfo {volume} {49}},\ \bibinfo {pages} {405--408}
  (\bibinfo {year} {1982})}\BibitemShut {NoStop}%
\bibitem [{\citenamefont {Ye}\ \emph {et~al.}(1999)\citenamefont {Ye},
  \citenamefont {Kim}, \citenamefont {Millis}, \citenamefont {Shraiman},
  \citenamefont {Majumdar},\ and\ \citenamefont {Te\ifmmode \check{s}\else
  \v{s}\fi{}anovi\ifmmode~\acute{c}\else \'{c}\fi{}}}]{ye1999}%
  \BibitemOpen
  \bibfield  {author} {\bibinfo {author} {\bibfnamefont {J.}~\bibnamefont
  {Ye}}, \bibinfo {author} {\bibfnamefont {Y.~B.}\ \bibnamefont {Kim}},
  \bibinfo {author} {\bibfnamefont {A.~J.}\ \bibnamefont {Millis}}, \bibinfo
  {author} {\bibfnamefont {B.~I.}\ \bibnamefont {Shraiman}}, \bibinfo {author}
  {\bibfnamefont {P.}~\bibnamefont {Majumdar}}, \ and\ \bibinfo {author}
  {\bibfnamefont {Z.}~\bibnamefont {Te\ifmmode \check{s}\else
  \v{s}\fi{}anovi\ifmmode~\acute{c}\else \'{c}\fi{}}},\ }\bibfield  {title}
  {\enquote {\bibinfo {title} {Berry phase theory of the anomalous {Hall}
  effect: Application to colossal magnetoresistance manganites},}\ }\href
  {\doibase 10.1103/PhysRevLett.83.3737} {\bibfield  {journal} {\bibinfo
  {journal} {Phys. Rev. Lett.}\ }\textbf {\bibinfo {volume} {83}},\ \bibinfo
  {pages} {3737--3740} (\bibinfo {year} {1999})}\BibitemShut {NoStop}%
\bibitem [{\citenamefont {Chen}\ \emph {et~al.}(2014)\citenamefont {Chen},
  \citenamefont {Niu},\ and\ \citenamefont {MacDonald}}]{chen2014}%
  \BibitemOpen
  \bibfield  {author} {\bibinfo {author} {\bibfnamefont {H.}~\bibnamefont
  {Chen}}, \bibinfo {author} {\bibfnamefont {Q.}~\bibnamefont {Niu}}, \ and\
  \bibinfo {author} {\bibfnamefont {A.~H.}\ \bibnamefont {MacDonald}},\
  }\bibfield  {title} {\enquote {\bibinfo {title} {Anomalous {Hall} effect
  arising from noncollinear antiferromagnetism},}\ }\href {\doibase
  10.1103/PhysRevLett.112.017205} {\bibfield  {journal} {\bibinfo  {journal}
  {Phys. Rev. Lett.}\ }\textbf {\bibinfo {volume} {112}},\ \bibinfo {pages}
  {017205} (\bibinfo {year} {2014})}\BibitemShut {NoStop}%
\bibitem [{\citenamefont {Zhang}\ \emph {et~al.}(2020)\citenamefont {Zhang},
  \citenamefont {Ishizuka}, \citenamefont {Zhang}, \citenamefont {Hal\'asz},\
  and\ \citenamefont {Batista}}]{zhang2020}%
  \BibitemOpen
  \bibfield  {author} {\bibinfo {author} {\bibfnamefont {S.-S.}\ \bibnamefont
  {Zhang}}, \bibinfo {author} {\bibfnamefont {H.}~\bibnamefont {Ishizuka}},
  \bibinfo {author} {\bibfnamefont {H.}~\bibnamefont {Zhang}}, \bibinfo
  {author} {\bibfnamefont {G.~B.}\ \bibnamefont {Hal\'asz}}, \ and\ \bibinfo
  {author} {\bibfnamefont {C.~D.}\ \bibnamefont {Batista}},\ }\bibfield
  {title} {\enquote {\bibinfo {title} {Real-space {Berry} curvature of
  itinerant electron systems with spin-orbit interaction},}\ }\href {\doibase
  10.1103/PhysRevB.101.024420} {\bibfield  {journal} {\bibinfo  {journal}
  {Phys. Rev. B}\ }\textbf {\bibinfo {volume} {101}},\ \bibinfo {pages}
  {024420} (\bibinfo {year} {2020})}\BibitemShut {NoStop}%
\bibitem [{\citenamefont {Li}\ \emph {et~al.}(2023)\citenamefont {Li},
  \citenamefont {Koo}, \citenamefont {Zhu}, \citenamefont {Behnia},\ and\
  \citenamefont {Yan}}]{li2023}%
  \BibitemOpen
  \bibfield  {author} {\bibinfo {author} {\bibfnamefont {X.}~\bibnamefont
  {Li}}, \bibinfo {author} {\bibfnamefont {J.}~\bibnamefont {Koo}}, \bibinfo
  {author} {\bibfnamefont {Z.}~\bibnamefont {Zhu}}, \bibinfo {author}
  {\bibfnamefont {K.}~\bibnamefont {Behnia}}, \ and\ \bibinfo {author}
  {\bibfnamefont {B.}~\bibnamefont {Yan}},\ }\bibfield  {title} {\enquote
  {\bibinfo {title} {Field-linear anomalous {Hall} effect and {Berry} curvature
  induced by spin chirality in the kagom\'e antiferromagnet {Mn}$_3${Sn}},}\
  }\href {\doibase 10.1038/s41467-023-37076-w} {\bibfield  {journal} {\bibinfo
  {journal} {Nature Communications}\ }\textbf {\bibinfo {volume} {14}},\
  \bibinfo {pages} {1642} (\bibinfo {year} {2023})}\BibitemShut {NoStop}%
\bibitem [{\citenamefont {Wickles}\ and\ \citenamefont
  {Belzig}(2013)}]{wickles2013}%
  \BibitemOpen
  \bibfield  {author} {\bibinfo {author} {\bibfnamefont {C}~\bibnamefont
  {Wickles}}\ and\ \bibinfo {author} {\bibfnamefont {W}~\bibnamefont
  {Belzig}},\ }\bibfield  {title} {\enquote {\bibinfo {title} {Effective
  quantum theories for {Bloch} dynamics in inhomogeneous systems with
  nontrivial band structure},}\ }\href {\doibase 10.1103/PhysRevB.88.045308}
  {\bibfield  {journal} {\bibinfo  {journal} {Phys. Rev. B}\ }\textbf {\bibinfo
  {volume} {88}},\ \bibinfo {pages} {045308} (\bibinfo {year}
  {2013})}\BibitemShut {NoStop}%
\bibitem [{\citenamefont {Mangeolle}\ \emph {et~al.}(2024)\citenamefont
  {Mangeolle}, \citenamefont {Savary},\ and\ \citenamefont
  {Balents}}]{mangeolle2024}%
  \BibitemOpen
  \bibfield  {author} {\bibinfo {author} {\bibfnamefont {L\'eo}\ \bibnamefont
  {Mangeolle}}, \bibinfo {author} {\bibfnamefont {Lucile}\ \bibnamefont
  {Savary}}, \ and\ \bibinfo {author} {\bibfnamefont {Leon}\ \bibnamefont
  {Balents}},\ }\bibfield  {title} {\enquote {\bibinfo {title} {Quantum kinetic
  equation and thermal conductivity tensor for bosons},}\ }\href {\doibase
  10.1103/PhysRevB.109.235137} {\bibfield  {journal} {\bibinfo  {journal}
  {Phys. Rev. B}\ }\textbf {\bibinfo {volume} {109}},\ \bibinfo {pages}
  {235137} (\bibinfo {year} {2024})}\BibitemShut {NoStop}%
\bibitem [{\citenamefont {Graf}\ and\ \citenamefont
  {Pi\'echon}(2021)}]{graf2021}%
  \BibitemOpen
  \bibfield  {author} {\bibinfo {author} {\bibfnamefont {A.}~\bibnamefont
  {Graf}}\ and\ \bibinfo {author} {\bibfnamefont {F.}~\bibnamefont
  {Pi\'echon}},\ }\bibfield  {title} {\enquote {\bibinfo {title} {Berry
  curvature and quantum metric in ${N}$-band systems: An eigenprojector
  approach},}\ }\href {\doibase 10.1103/PhysRevB.104.085114} {\bibfield
  {journal} {\bibinfo  {journal} {Phys. Rev. B}\ }\textbf {\bibinfo {volume}
  {104}},\ \bibinfo {pages} {085114} (\bibinfo {year} {2021})}\BibitemShut
  {NoStop}%
\bibitem [{\citenamefont {Graf}(2022)}]{grafthesis}%
  \BibitemOpen
  \bibfield  {author} {\bibinfo {author} {\bibfnamefont {A.}~\bibnamefont
  {Graf}},\ }\emph {\bibinfo {title} {Aspects of multiband systems: Quantum
  geometry, flat bands, and multifold fermions}},\ \href
  {https://theses.hal.science/tel-04047054} {\bibinfo {type} {Theses}},\
  \bibinfo  {school} {{Universit{\'e} Paris-Saclay}} (\bibinfo {year}
  {2022})\BibitemShut {NoStop}%
\bibitem [{\citenamefont {Provost}\ and\ \citenamefont
  {Vall\'ee}(1980)}]{provost1980}%
  \BibitemOpen
  \bibfield  {author} {\bibinfo {author} {\bibfnamefont {J.~P.}\ \bibnamefont
  {Provost}}\ and\ \bibinfo {author} {\bibfnamefont {G.}~\bibnamefont
  {Vall\'ee}},\ }\bibfield  {title} {\enquote {\bibinfo {title} {Riemannian
  structure on manifolds of quantum states},}\ }\href {\doibase
  10.1007/BF02193559} {\bibfield  {journal} {\bibinfo  {journal}
  {Communications in Mathematical Physics}\ }\textbf {\bibinfo {volume} {76}},\
  \bibinfo {pages} {289--301} (\bibinfo {year} {1980})}\BibitemShut {NoStop}%
\bibitem [{\citenamefont {Resta}(2011)}]{resta2011}%
  \BibitemOpen
  \bibfield  {author} {\bibinfo {author} {\bibfnamefont {R.}~\bibnamefont
  {Resta}},\ }\bibfield  {title} {\enquote {\bibinfo {title} {The insulating
  state of matter: a geometrical theory},}\ }\href {\doibase
  10.1140/epjb/e2010-10874-4} {\bibfield  {journal} {\bibinfo  {journal} {The
  European Physical Journal B}\ }\textbf {\bibinfo {volume} {79}},\ \bibinfo
  {pages} {121--137} (\bibinfo {year} {2011})}\BibitemShut {NoStop}%
\bibitem [{\citenamefont {Tatara}\ and\ \citenamefont
  {Kawamura}(2002)}]{tatara2002}%
  \BibitemOpen
  \bibfield  {author} {\bibinfo {author} {\bibfnamefont {G}~\bibnamefont
  {Tatara}}\ and\ \bibinfo {author} {\bibfnamefont {H}~\bibnamefont
  {Kawamura}},\ }\bibfield  {title} {\enquote {\bibinfo {title}
  {Chirality-driven anomalous {Hall} effect in weak coupling regime},}\ }\href
  {\doibase 10.1143/JPSJ.71.2613} {\bibfield  {journal} {\bibinfo  {journal}
  {Journal of the Physical Society of Japan}\ }\textbf {\bibinfo {volume}
  {71}},\ \bibinfo {pages} {2613--2616} (\bibinfo {year} {2002})},\ \Eprint
  {http://arxiv.org/abs/https://doi.org/10.1143/JPSJ.71.2613}
  {https://doi.org/10.1143/JPSJ.71.2613} \BibitemShut {NoStop}%
\bibitem [{\citenamefont {Fukui}\ \emph {et~al.}(2005)\citenamefont {Fukui},
  \citenamefont {Hatsugai},\ and\ \citenamefont {Suzuki}}]{fukui2005}%
  \BibitemOpen
  \bibfield  {author} {\bibinfo {author} {\bibfnamefont {Takahiro}\
  \bibnamefont {Fukui}}, \bibinfo {author} {\bibfnamefont {Yasuhiro}\
  \bibnamefont {Hatsugai}}, \ and\ \bibinfo {author} {\bibfnamefont {Hiroshi}\
  \bibnamefont {Suzuki}},\ }\bibfield  {title} {\enquote {\bibinfo {title}
  {Chern numbers in discretized {Brillouin} zone: Efficient method of computing
  (spin) {Hall} conductances},}\ }\href {\doibase 10.1143/JPSJ.74.1674}
  {\bibfield  {journal} {\bibinfo  {journal} {Journal of the Physical Society
  of Japan}\ }\textbf {\bibinfo {volume} {74}},\ \bibinfo {pages} {1674--1677}
  (\bibinfo {year} {2005})},\ \Eprint
  {http://arxiv.org/abs/https://doi.org/10.1143/JPSJ.74.1674}
  {https://doi.org/10.1143/JPSJ.74.1674} \BibitemShut {NoStop}%
\bibitem [{\citenamefont {Wei\ss{}e}\ \emph {et~al.}(2006)\citenamefont
  {Wei\ss{}e}, \citenamefont {Wellein}, \citenamefont {Alvermann},\ and\
  \citenamefont {Fehske}}]{weisse2006}%
  \BibitemOpen
  \bibfield  {author} {\bibinfo {author} {\bibfnamefont {A}~\bibnamefont
  {Wei\ss{}e}}, \bibinfo {author} {\bibfnamefont {G}~\bibnamefont {Wellein}},
  \bibinfo {author} {\bibfnamefont {A}~\bibnamefont {Alvermann}}, \ and\
  \bibinfo {author} {\bibfnamefont {H}~\bibnamefont {Fehske}},\ }\bibfield
  {title} {\enquote {\bibinfo {title} {The kernel polynomial method},}\ }\href
  {\doibase 10.1103/RevModPhys.78.275} {\bibfield  {journal} {\bibinfo
  {journal} {Rev. Mod. Phys.}\ }\textbf {\bibinfo {volume} {78}},\ \bibinfo
  {pages} {275--306} (\bibinfo {year} {2006})}\BibitemShut {NoStop}%
\bibitem [{\citenamefont {Ishizuka}\ and\ \citenamefont
  {Nagaosa}(2021)}]{ishizuka2021}%
  \BibitemOpen
  \bibfield  {author} {\bibinfo {author} {\bibfnamefont {Hiroaki}\ \bibnamefont
  {Ishizuka}}\ and\ \bibinfo {author} {\bibfnamefont {N}~\bibnamefont
  {Nagaosa}},\ }\bibfield  {title} {\enquote {\bibinfo {title} {Large anomalous
  {Hall} effect and spin {Hall} effect by spin-cluster scattering in the
  strong-coupling limit},}\ }\href {\doibase 10.1103/PhysRevB.103.235148}
  {\bibfield  {journal} {\bibinfo  {journal} {Phys. Rev. B}\ }\textbf {\bibinfo
  {volume} {103}},\ \bibinfo {pages} {235148} (\bibinfo {year}
  {2021})}\BibitemShut {NoStop}%
\bibitem [{\citenamefont {Nagaosa}\ \emph {et~al.}(2010)\citenamefont
  {Nagaosa}, \citenamefont {Sinova}, \citenamefont {Onoda}, \citenamefont
  {MacDonald},\ and\ \citenamefont {Ong}}]{nagaosa2010}%
  \BibitemOpen
  \bibfield  {author} {\bibinfo {author} {\bibfnamefont {N}~\bibnamefont
  {Nagaosa}}, \bibinfo {author} {\bibfnamefont {J}~\bibnamefont {Sinova}},
  \bibinfo {author} {\bibfnamefont {S}~\bibnamefont {Onoda}}, \bibinfo {author}
  {\bibfnamefont {A.~H.}\ \bibnamefont {MacDonald}}, \ and\ \bibinfo {author}
  {\bibfnamefont {N.~P.}\ \bibnamefont {Ong}},\ }\bibfield  {title} {\enquote
  {\bibinfo {title} {Anomalous {Hall} effect},}\ }\href {\doibase
  10.1103/RevModPhys.82.1539} {\bibfield  {journal} {\bibinfo  {journal} {Rev.
  Mod. Phys.}\ }\textbf {\bibinfo {volume} {82}},\ \bibinfo {pages}
  {1539--1592} (\bibinfo {year} {2010})}\BibitemShut {NoStop}%
\bibitem [{\citenamefont {Diop}\ \emph {et~al.}(2022)\citenamefont {Diop},
  \citenamefont {Jackeli},\ and\ \citenamefont {Savary}}]{diop2022}%
  \BibitemOpen
  \bibfield  {author} {\bibinfo {author} {\bibfnamefont {Seydou-Samba}\
  \bibnamefont {Diop}}, \bibinfo {author} {\bibfnamefont {George}\ \bibnamefont
  {Jackeli}}, \ and\ \bibinfo {author} {\bibfnamefont {Lucile}\ \bibnamefont
  {Savary}},\ }\bibfield  {title} {\enquote {\bibinfo {title} {Anisotropic
  exchange and noncollinear antiferromagnets on a noncentrosymmetric fcc
  half-{Heusler} structure},}\ }\href {\doibase 10.1103/PhysRevB.105.144431}
  {\bibfield  {journal} {\bibinfo  {journal} {Phys. Rev. B}\ }\textbf {\bibinfo
  {volume} {105}},\ \bibinfo {pages} {144431} (\bibinfo {year}
  {2022})}\BibitemShut {NoStop}%
\bibitem [{\citenamefont {Martin}\ and\ \citenamefont
  {Batista}(2008)}]{martin2008}%
  \BibitemOpen
  \bibfield  {author} {\bibinfo {author} {\bibfnamefont {Ivar}\ \bibnamefont
  {Martin}}\ and\ \bibinfo {author} {\bibfnamefont {C.~D.}\ \bibnamefont
  {Batista}},\ }\bibfield  {title} {\enquote {\bibinfo {title} {Itinerant
  electron-driven chiral magnetic ordering and spontaneous quantum {Hall}
  effect in triangular lattice models},}\ }\href {\doibase
  10.1103/PhysRevLett.101.156402} {\bibfield  {journal} {\bibinfo  {journal}
  {Phys. Rev. Lett.}\ }\textbf {\bibinfo {volume} {101}},\ \bibinfo {pages}
  {156402} (\bibinfo {year} {2008})}\BibitemShut {NoStop}%
\bibitem [{bel(2024)}]{bellwiki}%
  \BibitemOpen
  \href@noop {} {\enquote {\bibinfo {title} {{Bell} polynomials},}\ }\bibinfo
  {howpublished} {\url{https://en.wikipedia.org/wiki/Bell\_polynomials}}
  (\bibinfo {year} {2024}),\ \bibinfo {note} {[Online; accessed
  01-Nov-2024]}\BibitemShut {NoStop}%
\bibitem [{\citenamefont {Dittner}(1971)}]{dittner1971}%
  \BibitemOpen
  \bibfield  {author} {\bibinfo {author} {\bibfnamefont {P.}~\bibnamefont
  {Dittner}},\ }\bibfield  {title} {\enquote {\bibinfo {title} {Invariant
  tensors in ${SU}(3)$},}\ }\href {\doibase 10.1007/BF01877709} {\bibfield
  {journal} {\bibinfo  {journal} {Communications in Mathematical Physics}\
  }\textbf {\bibinfo {volume} {22}},\ \bibinfo {pages} {238 -- 252} (\bibinfo
  {year} {1971})}\BibitemShut {NoStop}%
\bibitem [{\citenamefont {Borodulin}\ \emph {et~al.}(2022)\citenamefont
  {Borodulin}, \citenamefont {Rogalyov},\ and\ \citenamefont
  {Slabospitskii}}]{borodulin2017}%
  \BibitemOpen
  \bibfield  {author} {\bibinfo {author} {\bibfnamefont {V.I.}\ \bibnamefont
  {Borodulin}}, \bibinfo {author} {\bibfnamefont {R.N.}\ \bibnamefont
  {Rogalyov}}, \ and\ \bibinfo {author} {\bibfnamefont {S.R.}\ \bibnamefont
  {Slabospitskii}},\ }\bibfield  {title} {\enquote {\bibinfo {title} {{CORE}
  3.2 ({COmpendium} of {RElations}, version 3.2)},}\ }\href
  {https://doi.org/10.48550/arXiv.1702.08246} {\bibfield  {journal} {\bibinfo
  {journal} {arXiv}\ ,\ \bibinfo {pages} {1702.08246v3}} (\bibinfo {year}
  {2022})}\BibitemShut {NoStop}%
\end{thebibliography}%

\newpage
\newpage 

\appendix

\section{Polynomials}
\label{sec:polynomials}

Here we extend the results of Refs.~\cite{graf2021,grafthesis} to the case of {\em traceful} SU(M) basis
matrices, and more specifically to the physical choice of basis we
make (Eq.~\eqref{eq:3}).

\subsection{Useful polynomial identities}
\label{sec:usef-polyn-ident}

Here we quote some well-known identities, also reviewed in
Ref.~\cite{graf2021,grafthesis}, 
useful for the derivations in Secs.~\ref{sec:proj-poly} and \ref{sec:berry-curv-terms}.

The `elementary symmetric polynomials' $\mathfrak{X}\mapsto\mathcal{E}_r(\mathfrak{X})$ with
$r=0,..,|\mathfrak{X}|$ ($|\mathfrak{X}|$ denotes the size of the set $\mathfrak{X}$) are the sums of all distinct
products of $r$ distinct variables taken from the set $\mathfrak{X}$, i.e.
\begin{align}
  \label{eq:28}
  &\mathcal{E}_0(\{x_1,\cdots,x_M\})=1,\qquad\mathcal{E}_1(\{x_1,\cdots,x_M\})=\sum_{i=1}^Mx_i,\nonumber\\
 & \mathcal{E}_2(\{x_1,\cdots,x_M\})=\sum_{1\leq i<j\leq
   M}x_ix_j,\qquad \cdots, \nonumber\\
  &
  \mathcal{E}_M(\{x_1,\cdots,x_M\})=\prod_{i=1}^Mx_i.
\end{align}
The `complete exponential Bell polynomials' $\mathfrak{X}\mapsto\mathcal{Y}_r(\mathfrak{X})$
with $r\in\mathbb{N}$ and $|\mathfrak{X}|=r$ are defined by
\begin{align}
  \label{eq:122}
  \mathcal{Y}_0(\{\})=1,\quad\mathcal{Y}_r(\{x_1,\cdots,x_r\})=r!\sum \prod_{i=1}^{r}\frac{x_i^{j_i}}{(i!)^{j_i}j_i!},
\end{align}
where the sum is taken over the $\{j_i\in\mathbb{N}\}_{i=1,..,r}$
such that
$\sum_{i=1}^rij_i=r$ \cite{bellwiki}. 

Now we define the morphism of sets $\mathfrak{X}\mapsto \mathfrak{S}_r(\mathfrak{X})$
\begin{align}
  \label{eq:123}
  \mathfrak{S}_0(\mathfrak{X})&=\{\},\\
  \mathfrak{S}_r(\{x_1,..,x_p\})&=\left\{(-1)^{k-1}(k-1)!\left(\sum_{i=1}^px_i ^k\right)\right\}_{k=1,..,r}\nonumber.
\end{align}
Note that $|\mathfrak{S}_r(\mathfrak{X})|=r$ independently of $|\mathfrak{X}|$. Newton's identities allow to show that
\begin{equation}
  \label{eq:124}
  \mathcal{E}_r(\mathfrak{X})=\frac{1}{r!}\mathcal{Y}_r(\mathfrak{S}_r(\mathfrak{X})).
\end{equation}

These relations are particularly useful when expressing polynomials of
the form:
\begin{align}
  \label{eq:32}
  &\prod_{m=1}^M(x-x_m)\\
  &=(-1)^M\prod_{m=1}^Mx_m+\cdots-x^{M-1}\sum_{m=1}^Mx_m+x^M\nonumber\\
  &=\sum_{r=0}^M(-1)^{M-r}\mathcal{E}_{M-r}(\{x_1,\cdots,x_M\})x^{r}\nonumber\\
  &=\sum_{r=0}^M\frac{(-1)^{M-r}}{(M-r)!}\mathcal{Y}_{M-r}\Big(\mathfrak{S}_{M-r}(\{x_1,\cdots,x_M\})\Big)x^{r}.
    \nonumber
\end{align}

\subsection{Projector as a polynomial in $H$}
\label{sec:proj-poly}

We have, in general, away from degeneracies, for
$\{\varepsilon_m\}_{m=1,..,M}$ the eigenvalues of a $M\times M$ Hamiltonian $\hat{H}$, and
$\hat{P}_{(n)}$ the projector onto the $n$th band of $\hat{H}$,
\begin{equation}
  \label{eq:14}
  \hat{P}_{(n)}=\prod_{m\neq n}\frac{\hat{H}-\varepsilon_{m}}{\varepsilon_{n}-\varepsilon_{m}}=\frac{\prod_{m\neq n}(\hat{H}-\varepsilon_{m})}{\prod_{m\neq n}(\varepsilon_{n}-\varepsilon_{m})}.
\end{equation}
If we expand the products in the numerator, we obtain
\begin{align}
  \label{eq:21}
  &\prod_{m\neq n}(\hat{H}-\varepsilon_{m})\\
  &=\sum_{r=0}^{M-1}\frac{(-1)^{(M-1)-r}}{[(M-1)-r]!}\mathcal{Y}_{(M-1)-r}\Big(\mathfrak{S}_{(M-1)-r}^{(n)}(\hat{H})\Big)\hat{H}^{r},\nonumber
\end{align}
where
\begin{align}
  \label{eq:125}
  \mathfrak{S}_0^{(n)}(\hat{H})&\equiv\{\},\\
  \mathfrak{S}_{r>0}^{(n)}(\hat{H})&\equiv \mathfrak{S}_r(\{\varepsilon_1,\cdots,\varepsilon_{n-1},\varepsilon_{n+1},\cdots,\varepsilon_M\})\nonumber\\
  &=\left\{(-1)^{k-1}(k-1)!\left({\rm
        Tr}\hat{H}^k-\varepsilon_n^k\right)\right\}_{k=1,..,r}.\nonumber
\end{align}
Similarly,
\begin{align}
  \label{eq:126}
&\prod_{m\neq n}(\varepsilon_n-\varepsilon_{m})\nonumber\\
  &=\sum_{r=0}^{M-1}\frac{(-1)^{(M-1)-r}}{[(M-1)-r]!}\mathcal{Y}_{(M-1)-r}\Big(\mathfrak{S}_{(M-1)-r}^{(n)}(\hat{H})\Big)\varepsilon_n^{r}\nonumber,\\
  &\equiv \mathcal{N}_{(n)}(\hat{H})
\end{align}
and we define, for $r=0,..,M-1$,
\begin{equation}
  \label{eq:127}
  \ell^{(n)}_{r}(\hat{H})\equiv \frac{(-1)^{(M-1)-r}}{[(M-1)-r]!}\frac{\mathcal{Y}_{(M-1)-r}\Big(\mathfrak{S}_{(M-1)-r}^{(n)}(\hat{H})\Big)
  }{\mathcal{N}_{(n)}(\hat{H})},
\end{equation}
so that
\begin{equation}
  \label{eq:128}
  \hat{P}_{(n)}=\sum_{r=0}^{M-1} \ell^{(n)}_{r}\hat{H}^r,
\end{equation}
as in the main text, Eq.~\eqref{eq:101}.

\subsection{Expressions for $M=6$}
\label{sec:expressions-m=6}

Here we give the expressions for $\mathcal{N}_{(n)}$ and
$\ell_r^{(n)}$ defined in Eqs.~(\ref{eq:126},\ref{eq:127}) for a $6\times6$
Hamiltonian. Defining, $\forall r\in\mathbb{N}$, $\mathcal{C}_r(\bs{k})\equiv{\rm
  Tr}\hat{H}^r(\bs{k})=\sum_{n=1}^{M}\varepsilon_n^r(\bs{k})$, i.e.\ the
`$r$ power sum' of the eigenvalues of $\hat{H}$ and
\begin{align}
  \label{eq:43}
  \mathcal{N}_{(n)}&=6\varepsilon_n^5-5\mathcal{C}_1\varepsilon_n^4+2(\mathcal{C}_1^2-\mathcal{C}_2)\varepsilon_n^3\\
  &\quad+\frac{1}{2}\left(-\mathcal{C}_1^3+3\mathcal{C}_1\mathcal{C}_2-2\mathcal{C}_3\right)\varepsilon_n^2\nonumber\\
  &\quad+\frac{1}{12}\left(\mathcal{C}_1^4-6\mathcal{C}_1^2\mathcal{C}_2+3\mathcal{C}_2^2+8\mathcal{C}_1\mathcal{C}_3-6\mathcal{C}_4\right)\varepsilon_n\nonumber\\
             &\quad+\frac{1}{120}\left(-\mathcal{C}_1^5+10\mathcal{C}_1^3\mathcal{C}_2-20\mathcal{C}_1^2\mathcal{C}_3+20\mathcal{C}_2\mathcal{C}_3\right.\nonumber\\
  &\quad\qquad\qquad\left.-15\mathcal{C}_1(\mathcal{C}_2^2-2\mathcal{C}_4)-24\mathcal{C}_5\right),\nonumber
\end{align}
and
\begin{align}
  \label{eq:44}
  \mathcal{N}_{(n)}\ell_0^{(n)}&=\varepsilon_n^5-\mathcal{C}_1\varepsilon_n^4+\frac{1}{2}(\mathcal{C}_1^2-\mathcal{C}_2)\varepsilon_n^3\nonumber\\
        &\quad+\frac{1}{6}(-\mathcal{C}_1^3+3\mathcal{C}_1\mathcal{C}_2-2\mathcal{C}_3)\varepsilon_n^2\nonumber\\
       &\quad+\frac{1}{24}(\mathcal{C}_1^4-6\mathcal{C}_1^2\mathcal{C}_2+3\mathcal{C}_2^2+8\mathcal{C}_1\mathcal{C}_3-6\mathcal{C}_4)\varepsilon_n\nonumber\\
                               &\quad+\frac{1}{120}\left(-\mathcal{C}_1^5+10\mathcal{C}_1^3\mathcal{C}_2-20\mathcal{C}_1^2\mathcal{C}_3+20\mathcal{C}_2\mathcal{C}_3\right.\nonumber\\
  &\quad\qquad\qquad\left.-15\mathcal{C}_1(\mathcal{C}_2^2-2\mathcal{C}_4)-24\mathcal{C}_5\right),\nonumber\\
  \mathcal{N}_{(n)}\ell_1^{(n)}&=\varepsilon_n^4-\mathcal{C}_1\varepsilon_n^3+\frac{1}{2}(\mathcal{C}_1^2-\mathcal{C}_2)\varepsilon_n^2\nonumber\\
        &\quad+\frac{1}{6}(-\mathcal{C}_1^3+3\mathcal{C}_1\mathcal{C}_2-2\mathcal{C}_3)\varepsilon_n\nonumber\\
  &\quad+\frac{1}{24}(\mathcal{C}_1^4-6\mathcal{C}_1^2\mathcal{C}_2+3\mathcal{C}_2^2+8\mathcal{C}_1\mathcal{C}_3-6\mathcal{C}_4),\nonumber\\
  \mathcal{N}_{(n)}\ell_2^{(n)}&=\varepsilon_n^3-\mathcal{C}_1\varepsilon_n^2+\frac{1}{2}(\mathcal{C}_1^2-\mathcal{C}_2)\varepsilon_n\nonumber\\
  &\quad+\frac{1}{6}(-\mathcal{C}_1^3+3\mathcal{C}_1\mathcal{C}_2-2\mathcal{C}_3),\nonumber\\
  \mathcal{N}_{(n)}\ell_3^{(n)}&=\varepsilon_n^2-\mathcal{C}_1\varepsilon_n+\frac{1}{2}(\mathcal{C}_1^2-\mathcal{C}_2),\nonumber\\
  \mathcal{N}_{(n)}\ell_4^{(n)}&=\varepsilon_n-\mathcal{C}_1,\nonumber\\
  \mathcal{N}_{(n)}\ell_5^{(n)}&=1.
\end{align}

\section{Berry curvature in terms of powers of the Hamiltonian and
  eigenenergies}
\label{sec:berry-curv-terms}

\subsection{Expressions for the quantum geometric tensor}
\label{sec:expr-quant-metr}

We start by recalling the definition of the quantum geometric tensor in
terms of projection operators into bands \cite{resta2011,graf2021},
\begin{align}
  \label{eq:6}
  G_{(n)}^{\alpha\beta}
&={\rm Tr}[\partial_\alpha \hat{P}_{(n)}(1-\hat{P}_{(n)})\partial_\beta \hat{P}_{(n)}].
\end{align}
Its symmetric part under $\alpha\leftrightarrow\beta$ is the quantum
metric
\begin{align}
  \label{eq:38}
  \Gamma^{\alpha\beta}_{(n)}&=\frac{1}{2}(G_{(n)}^{\alpha\beta}+G_{(n)}^{\beta\alpha})\nonumber\\      
  &={\rm Tr}[\partial_\alpha \hat{P}_{(n)}\partial_\beta
    \hat{P}_{(n)}]-\frac{1}{2}{\rm Tr}[\{\partial_\beta
    \hat{P}_{(n)},\partial_\alpha
    \hat{P}_{(n)}\}\hat{P}_{(n)}]\nonumber\\
  &=\frac{1}{2}(G_{(n)}^{\alpha\beta}+[G_{(n)}^{\alpha\beta}]^*)\nonumber\\
  &={\rm Re}G^{\alpha\beta}_{(n)},
\end{align}
and the Berry curvature is
\begin{align}
  \label{eq:39}
  \Omega^{\alpha\beta}_{(n)}&=i(G_{(n)}^{\alpha\beta}-G_{(n)}^{\beta\alpha})\nonumber\\
  &=-i{\rm Tr}[[\partial_\beta
    \hat{P}_{(n)},\partial_\alpha
    \hat{P}_{(n)}]\hat{P}_{(n)}]\nonumber\\
  &={\rm Im}\left({\rm Tr}[[\partial_\beta
    \hat{P}_{(n)},\partial_\alpha
    \hat{P}_{(n)}]\hat{P}_{(n)}]\right)\nonumber\\
  &=i(G_{(n)}^{\alpha\beta}-[G_{(n)}^{\alpha\beta}]^*)\nonumber\\
  &=-2{\rm Im}G^{\alpha\beta}_{(n)}\nonumber\\
  &=2{\rm Im}{\rm Tr}[\partial_\alpha\hat{P}_{(n)}\hat{P}_{(n)}\partial_\beta\hat{P}_{(n)}],
\end{align}
since ${\rm Im}{\rm Tr}[\partial_\alpha\hat{P}_{(n)}\partial_\beta\hat{P}_{(n)}]=0$.

In the more conventional form of eigenvectors,
\begin{align}
  \label{eq:29}
  G_{(n)}^{\alpha\beta}&=\langle\partial_\alpha\psi_n|(1-\hat{P}_{(n)})|\partial_\beta\psi_n\rangle\nonumber\\
                       &=\sum_{m\neq n}\frac{\langle \psi_n|\partial_\alpha
  \hat{H}|\psi_m\rangle\langle \psi_m|\partial_\beta
  \hat{H}|\psi_n\rangle}{[\varepsilon_n-\varepsilon_m]^2},
\end{align}
and
\begin{align}
  \label{eq:30}
  \Gamma^{\alpha\beta}_{(n)}={\rm Re}\langle\partial_\alpha\psi_n|(1-\hat{P}_{(n)})|\partial_\beta\psi_n\rangle,
\end{align}
\begin{align}
  \label{eq:31}
  \Omega^{\alpha\beta}_{(n)}&=-2{\rm
                              Im}\langle\partial_\alpha\psi_n|(1-\hat{P}_{(n)})|\partial_\beta\psi_n\rangle\nonumber\\
  &=-2{\rm
    Im}\langle\partial_\alpha\psi_n|\partial_\beta\psi_n\rangle\nonumber\\
  &=i(\partial_\alpha[\langle\psi_n|\partial_\beta\psi_n\rangle]-\partial_\beta[\langle\psi_n|\partial_\alpha\psi_n\rangle]).
\end{align}

\subsection{In terms of powers of the Hamiltonian and eigenenergies}
\label{sec:terms-powers-hamilt-1}

In turn, using Eq.~\eqref{eq:128},
\begin{widetext}
\begin{align}
  \label{eq:143}
  \Omega^{\alpha\beta}_{(n)}&=2\sum_{r_1,r_2,r_3=0}^{M-1}{\rm Im}{\rm Tr}\left[\partial_\alpha[\ell_{r_1}^{(n)}
                         \hat{H}^{r_1}]\ell_{r_2}^{(n)}\hat{H}^{r_2}\partial_\beta[
                         \ell_{r_3}^{(n)}\hat{H}^{r_3}]\right],\nonumber\\
  \Gamma_{(n)}^{\alpha\beta}&=\sum_{r_1,r_2=0}^{M-1}{\rm
                              Tr}\left[\partial_\alpha
                              [\ell_{r_1}^{(n)}\hat{H}^{r_1}]\partial_\beta
                              [\ell_{r_2}^{(n)}\hat{H}^{r_2}]\right]-\sum_{r_1,r_2,r_3=0}^{M-1}{\rm
                              Re}{\rm Tr}\left[\partial_\alpha[\ell_{r_1}^{(n)}
                         \hat{H}^{r_1}]\ell_{r_2}^{(n)}\hat{H}^{r_2}\partial_\beta[
                         \ell_{r_3}^{(n)}\hat{H}^{r_3}]\right].
\end{align}
Note that this in principle requires to differentiate the
$\ell$'s. However, as noted in Ref.~\cite{graf2021,grafthesis}, an
important simplification arises in the case of the Berry curvature,
which we now examine. Using the chain rule, we have
\begin{align}
  \label{eq:1}
 \Omega^{\alpha\beta}_{(n)}&=2\sum_{r_1,r_2,r_3=0}^{M-1}\ell_{r_1}^{(n)}\ell_{r_2}^{(n)}
                         \ell_{r_3}^{(n)}
                         {\rm Im}{\rm Tr}\left[\partial_\alpha[\hat{H}^{r_1}]\hat{H}^{r_2}\partial_\beta[\hat{H}^{r_3}]\right]+2\sum_{r_1,r_2,r_3=0}^{M-1}\partial_\alpha\ell_{r_1}^{(n)}\ell_{r_2}^{(n)}\partial_\beta
                         \ell_{r_3}^{(n)}{\rm Im}{\rm Tr}\left[
                         \hat{H}^{r_1}\hat{H}^{r_2}\hat{H}^{r_3}\right]\nonumber\\
  &\quad+2\sum_{r_1,r_2,r_3=0}^{M-1}\ell_{r_1}^{(n)}\ell_{r_2}^{(n)}\partial_\beta
                         \ell_{r_3}^{(n)}{\rm Im}{\rm Tr}\left[\partial_\alpha[\hat{H}^{r_1}]\hat{H}^{r_2}\hat{H}^{r_3}\right]+2\sum_{r_1,r_2,r_3=0}^{M-1}\partial_\alpha\ell_{r_1}^{(n)}\ell_{r_2}^{(n)}
                         \ell_{r_3}^{(n)}
                         {\rm Im}{\rm Tr}\left[\hat{H}^{r_1}\hat{H}^{r_2}\partial_\beta[\hat{H}^{r_3}]\right].
\end{align}
Using the antisymmetry of $\Omega^{\alpha\beta}_{(n)}$ under the
$\alpha\leftrightarrow\beta$ exchange, i.e.\ $\Omega^{\beta\alpha}_{(n)}=-\Omega^{\alpha\beta}_{(n)}$, we now show that only the first
term of Eq.~\eqref{eq:1} survives. Indeed, relabeling the dummy
indices $r_1\leftrightarrow r_3$ in half the terms, we find
\begin{align}
  \label{eq:27}
  \frac{1}{2}(\Omega^{\alpha\beta}_{(n)}-\Omega^{\beta\alpha}_{(n)})&=\sum_{r_1,r_2,r_3=0}^{M-1}\ell_{r_1}^{(n)}\ell_{r_2}^{(n)}
                         \ell_{r_3}^{(n)}
                         {\rm Im}{\rm Tr}\left[[\partial_\beta[\hat{H}^{r_3}],\partial_\alpha[\hat{H}^{r_1}]]\hat{H}^{r_2}\right]+\sum_{r_1,r_2,r_3=0}^{M-1}\partial_\alpha\ell_{r_1}^{(n)}\ell_{r_2}^{(n)}\partial_\beta
                         \ell_{r_3}^{(n)}{\rm Im}{\rm Tr}\left[
                         [\hat{H}^{r_3},\hat{H}^{r_1}]\hat{H}^{r_2}\right]\nonumber\\
  &\quad+\sum_{r_1,r_2,r_3=0}^{M-1}\ell_{r_1}^{(n)}\ell_{r_2}^{(n)}\partial_\beta
                         \ell_{r_3}^{(n)}{\rm Im}{\rm Tr}\left[\partial_\alpha[\hat{H}^{r_1}][\hat{H}^{r_2},\hat{H}^{r_3}]\right]+\sum_{r_1,r_2,r_3=0}^{M-1}\partial_\alpha\ell_{r_1}^{(n)}\ell_{r_2}^{(n)}
                         \ell_{r_3}^{(n)}
                         {\rm Im}{\rm Tr}\left[[\hat{H}^{r_1},\hat{H}^{r_2}]\partial_\beta[\hat{H}^{r_3}]\right].
\end{align}
Since
$[\hat{H}^{r},\hat{H}^{r'}]=0$, we have 
\begin{align}
  \label{eq:22}
 \Omega^{\alpha\beta}_{(n)}&=2\sum_{r_1,r_2,r_3=1}^{M-1}\ell^{(n)}_{r_1}\ell_{r_2}^{(n)}
                         \ell_{r_3}^{(n)}
                         {\rm Im}{\rm Tr}\left[\partial_\alpha[\hat{H}^{
                         r_1}]\hat{H}^{r_2}\partial_\beta[\hat{H}^{r_3}]\right],
\end{align}
where we removed the $r_i=0$ terms in the sum since they identically
vanish.  Indeed, $\hat{H}^{0}\equiv{\rm Id}_M$ has only constant elements, so
its derivative vanishes. Moreover, note that we can rewrite the $r_2=0$ contribution
as $\partial_\alpha[\hat{H}^{r_1}]\hat{H}^{0}
\partial_\beta[\hat{H}^{r_3}]$ whose sum over $r_1,r_3$ in
$\Omega^{\alpha\beta}$ vanishes since it is symmetric under
$\alpha\leftrightarrow\beta$ while $\Omega^{\alpha\beta}$ is
antisymmetric under this exchange.

\end{widetext}

\section{Explicit contractions in spin space}
\label{sec:contractions-spin-space}

\subsection{General relations}
\label{sec:general-relations}

In sublattice space,
\begin{equation}
  \label{eq:35}
 \hat{E}_{a_1a_2}\hat{E}_{a_3a_4}\cdots\hat{E}_{a_{2r-1}a_{2r}}=\delta_{a_2a_3}\delta_{a_4a_5}\cdots \delta_{a_{2r-2}a_{2r-1}}\hat{E}_{a_1a_{2r}}
\end{equation}
and, in spin space, for $r\geq2$,
\begin{equation}
  \label{eq:500}
  \hat{\sigma}_{\mu_1}\hat{\sigma}_{\mu_2}\cdots\hat{\sigma}_{\mu_r}=\sum_{\{\rho_i\}_{i=2,..,r}}\!\!\!\!g_{\mu_1\mu_2\rho_2}g_{\rho_2\mu_3\rho_3}\cdots g_{\rho_{r-1}\mu_r\rho_{r}}\hat{\sigma}_{\rho_r}.
\end{equation}

We have for $\mu,\nu,\rho=0,1,2,3$, $\zeta=\pm1$ and
$[\hat{A},\hat{B}]_\zeta=\hat{A}\hat{B}+\zeta\hat{B}\hat{A}$ for two
matrices $\hat{A}$ and $\hat{B}$:
\begin{align}
  \label{eq:109}
  &{\rm Tr}\left([\hat{\sigma}^\mu,
    \hat{\sigma}^\nu]_\zeta\hat{\sigma}^\rho\right)\\
  &=4i\frac{(1-\zeta)}{2}\epsilon_{\mu\nu\rho}+4 \frac{(1+\zeta)}{2}\left(\delta_{(\mu\nu}\delta_{0\rho)}-2\delta_{0\mu}\delta_{0\nu}\delta_{0\rho}\right),\nonumber
\end{align}
where $\delta_{(\mu\nu}\delta_{0\rho)}\equiv
\delta_{\mu\nu}\delta_{0\rho}+\delta_{\rho\mu}\delta_{0\nu}+\delta_{\nu\rho}\delta_{0\mu}$,
and $\epsilon_{\mu\nu\rho}$ is an abuse of notation for the 3d
Levi-Civita tensor such that
that $\epsilon_{\mu\nu\rho}=0$ if any of the $\mu,\nu,\rho=0$, and
\begin{align}
  \label{eq:40}
 g_{\mu\nu\rho} &\equiv\frac{1}{2}{\rm Tr}\left(\hat{\sigma}^\mu
    \hat{\sigma}^\nu\hat{\sigma}^\rho\right)\nonumber\\
  &=i\epsilon_{\mu\nu\rho}+\left(\delta_{(\mu\nu}\delta_{0\rho)}-2\delta_{0\mu}\delta_{0\nu}\delta_{0\rho}\right).
\end{align}
We note the identity (excluding $\mu,\nu,\rho,\lambda,\kappa,\tau=0$)
\begin{align}
  \label{eq:4}
  \epsilon_{\mu\nu\rho}\epsilon_{\lambda\kappa\tau}&=\delta_{\mu\tau}(\delta_{\nu\lambda}\delta_{\rho\kappa}-\delta_{\nu\kappa}\delta_{\rho\lambda})\nonumber\\
   &\quad+\delta_{\mu\lambda}(-\delta_{\nu\tau}\delta_{\rho\kappa}+\delta_{\nu\kappa}\delta_{\rho\tau})\nonumber\\
  &\quad-\delta_{\mu\kappa}(-\delta_{\nu\tau}\delta_{\rho\lambda}+\delta_{\nu\lambda}\delta_{\rho\tau}),
\end{align}
and in particular,
\begin{align}
  \label{eq:5}
  \epsilon_{\mu\nu\rho}\epsilon_{\rho\kappa\tau}=-\delta_{\mu\tau}\delta_{\nu\kappa}+\delta_{\mu\kappa}\delta_{\nu\tau}.
\end{align}


We have the following relations
\begin{align}
  \label{eq:26}
  {\rm
  Tr}[\hat{\sigma}_{\mu_1}]&=2g_{\mu_100}=2\delta_{\mu_10},\nonumber\\
  {\rm
  Tr}[\hat{\sigma}_{\mu_1}\hat{\sigma}_{\mu_2}]&=2g_{\mu_1\mu_20}=2\delta_{\mu_1\mu_2},\nonumber\\
  {\rm
  Tr}[\hat{\sigma}_{\mu_1}\hat{\sigma}_{\mu_2}\hat{\sigma}_{\mu_3}]&=2g_{\mu_1\mu_2\mu_3},
\end{align}
as well as
\begin{equation}
  \label{eq:11}
  g_{\mu\nu\rho}^*=g_{\rho\nu\mu},
\end{equation}
and for $r>3$, using Eq.~\eqref{eq:500},
\begin{align}
  \label{eq:36}
  {\rm
  Tr}[\hat{\sigma}_{\mu_1}\hat{\sigma}_{\mu_2}\cdots\hat{\sigma}_{\mu_r}]&=\sum_{\{\rho_i\}_{i=2,..,r}}g_{\mu_1\mu_2\rho_2}\cdots
                                                                            g_{\rho_{r-1}\mu_r\rho_r}{\rm
                                                                           Tr}\hat{\sigma}_{\rho_r}\nonumber\\
  &=2\!\!\!\!\sum_{\{\rho_i\}_{i=2,..,r}}g_{\mu_1\mu_2\rho_2}\cdots g_{\rho_{r-1}\mu_r\rho_r}\delta_{\rho_r0}\nonumber\\
  &=2\!\!\!\!\sum_{\{\rho_i\}_{i=2,..,r-1}}\!\!\!\!g_{\mu_1\mu_2\rho_2}g_{\rho_2\mu_3\rho_3}\cdots
                                                                           g_{\rho_{r-1}\mu_r0}\nonumber\\
  &=2\!\!\!\!\sum_{\{\rho_i\}_{i=2,..,r-2}}\!\!\!\!g_{\mu_1\mu_2\rho_2}g_{\rho_2\mu_3\rho_3}\cdots
    g_{\rho_{r-2}\mu_{r-1}\mu_{r}}.
\end{align}
Now, since, for
any $\mu_0$, $\hat{\sigma}^{\mu_0}\hat{\sigma}^{\mu_0}={\rm Id}_2$ (no
summation), we have also
$\sum_{\mu_0=0}^4\hat{\sigma}^{\mu_0}\hat{\sigma}^{\mu_0}=4{\rm
  Id}_2$, and
\begin{align}
  \label{eq:46}
  & {\rm
    Tr}[\hat{\sigma}_{\mu_1}\hat{\sigma}_{\mu_2}\cdots\hat{\sigma}_{\mu_r}]\nonumber\\
  &\qquad=\frac{1}{4}\sum_{\mu_0}{\rm
                                                                           Tr}[\hat{\sigma}_{\mu_0}\hat{\sigma}_{\mu_1}\hat{\sigma}_{\mu_2}\cdots\hat{\sigma}_{\mu_r}\hat{\sigma}_{\mu_0}]\nonumber\\
  &\qquad=\frac{1}{2}\!\!\!\!\sum_{\mu_0,\{\rho_i\}_{i=1,..,r+1}}g_{\mu_0\mu_1\rho_1}\cdots
    g_{\rho_{r-1}\mu_r\rho_r}g_{\rho_{r}\mu_0\rho_{r+1}}\delta_{\rho_{r+1}0}\nonumber\\
  &\qquad=\frac{1}{2}\!\!\!\!\sum_{\mu_0,\{\rho_i\}_{i=1,..,r}}g_{\mu_0\mu_1\rho_1}\cdots
    g_{\rho_{r-1}\mu_r\rho_r}g_{\rho_{r}\mu_00}\nonumber\\
  &\qquad=\frac{1}{2}\!\!\!\!\sum_{\{\rho_i\}_{i=1,..,r}}g_{\rho_r\mu_1\rho_1}\cdots g_{\rho_{r-1}\mu_r\rho_r}.
\end{align}

\begin{widetext}

\subsection{Explicit traces of $\hat{\sigma}^{x,y,z}$ products}
\label{sec:x-y-z}

In this section, $\mu_i=x,y,z=1,2,3$, i.e.\ we exclude $\mu_i=0$, and
we provide the traces of the products of up to 5 Pauli matrices (the
expressions for a larger number of Pauli matrices become very long and
we deemed it not very instructive or useful to write them):
\begin{align}
  \label{eq:45}
  & {\rm Tr}[\sigma^{\mu_1}]=0,\\
 & {\rm Tr}[\sigma^{\mu_1}\sigma^{\mu_2}]=2\delta _{\mu _1\mu
   _2},\nonumber\\
 & {\rm Tr}[\sigma^{\mu_1}\sigma^{\mu_2}\sigma^{\mu_3}]=2i
   \epsilon_{\mu _1\mu _2\mu _3},\nonumber\\
 & {\rm Tr}[\sigma^{\mu_1}\sigma^{\mu_2}\sigma^{\mu_3}\sigma^{\mu_4}]=2(\delta_{\mu _1\mu _4} \delta_{\mu _2\mu
   _3}-\delta_{\mu _1\mu _3} \delta_{\mu _2\mu
   _4}+\delta_{\mu_1\mu_2}\delta_{\mu_3\mu_4}),\nonumber\\
 & {\rm Tr}[\sigma^{\mu_1}\sigma^{\mu_2}\sigma^{\mu_3}\sigma^{\mu_4}\sigma^{\mu_5}]=2i(\delta_{\mu _2\mu _3} \epsilon_{\mu _1\mu _4\mu
   _5}-\delta_{\mu _1\mu _3} \epsilon_{\mu
   _2\mu _4\mu _5}+\delta_{\mu_4\mu_5}\epsilon_{\mu_1\mu_2\mu_3}+\delta_{\mu_1\mu_2}\epsilon_{\mu_3\mu_4\mu_5}).\nonumber
\end{align}
Note that cyclic permutation of the indices are identical because of
the cyclicity of the trace, and that traces of an odd number of Pauli
matrices are purely imaginary while those of an even number of Pauli
matrices are purely real. For our application to three-sublattice
systems, it is also important that, up to thirteen Pauli matrices, the
traces may always be reduced to a form where either zero (even number
of matrices in the trace) or only one (odd number of matrices in the
trace) Levi-Civita symbol appears. The consequence is that only a
single power of the chirality within the unit cell, $\chi_{123}$, can
appear, cf.\ Eq.~\eqref{eq:24}, where $i_{123}=1$. Finally note
the interesting results in Refs.~\cite{dittner1971,borodulin2017}.

\newpage

\end{widetext}

\section{Decomposition of the Hamiltonian into tensor product bases}
\label{sec:deriv-analyt-expr}

We have, for $a,b=1,..,N$,
\begin{equation}
  \label{eq:174}
    \hat{H}=\left.\begin{pmatrix}
    \cdots & \cdots &\cdots & \cdots & \cdots \\
    \cdots & \hat{H}_{aa} & \cdots & \hat{H}_{a<b} & \cdots \\
    \cdots & \cdots & \cdots & \cdots &\cdots \\
    \cdots & \hat{H}_{b>a} & \cdots & \cdots & \cdots \\
    \cdots & \cdots & \cdots & \cdots & \cdots \\
    \end{pmatrix}\right\}M=2N,
\end{equation}
with $\hat{H}_{ab}$ $2\times2$ matrices:
\begin{align}
  \label{eq:176}
  \hat{H}_{ab}
  &=\sum_{\mu=0}^3 {\sf h}_{ab}^\mu\hat{\sigma}^\mu,
\end{align}
such that $\hat{H}_{ba}^{\vphantom{\dagger}}=\hat{H}_{ab}^\dagger$
since $\hat{H}$ is hermitian, and so $({\sf h}_{ab}^\mu)^*={\sf h}_{ba}^\mu$. Note that $\hat{H}_{ab}$ itself is not
necessarily hermitian (in turn, the ${\sf h}_{ab}^\mu$ can be
complex). We may also write
\begin{align}
  \label{eq:25}
  \hat{H}&=\sum_{\mu=0}^3\hat{\sf H}_\mu\otimes\hat{\sigma}^\mu,
\end{align}
where
\begin{equation}
  \label{eq:12}
 \hat{\sf H}_\mu= \underbrace{\begin{pmatrix}
    \cdots & \cdots &\cdots & \cdots & \cdots \\
    \cdots & {\sf h}^\mu_{aa} & \cdots & {\sf h}_{a<b}^\mu & \cdots \\
    \cdots & \cdots & \cdots & \cdots &\cdots \\
    \cdots & {\sf h}^\mu_{b>a} & \cdots & \cdots & \cdots \\
    \cdots & \cdots & \cdots & \cdots & \cdots \\
    \end{pmatrix}}_N,
\end{equation}
where $\hat{\sf H}_\mu^\dagger=\hat{\sf H}_\mu$.


\section{Three-sublattice structure}
\label{sec:three-subl-struct}

Recall that we defined
\begin{equation}
  \label{eq:47}
  \Lambda_{(q_1,q_2)}^{\alpha\beta}\equiv{\rm Tr}[\partial_\alpha\hat{H}\hat{H}^{q_1}\partial_\beta\hat{H}\hat{H}^{q_2}],
\end{equation}
and we have
\begin{equation}
  \label{eq:48}
  \Omega_{(n)}^{\alpha\beta}=2\sum_{r_{1,2,3}=1}^{M-1}\ell_{r_1}^{(n)}\ell_{r_2}^{(n)}\ell_{r_3}^{(n)}\sum_{p=r_2+2}^{R}\xi_{\{r_i\}}(p){\rm
  Im}\Lambda_{(p-2,R-p)}^{\alpha\beta}.
\end{equation}
For a
given set of values $(r_1,r_2,r_3)$ and their permutations, all terms $\Lambda^{\alpha\beta}_{(q_1,q_2)}\equiv {\rm
  Tr}[\partial_\alpha\hat{H}\hat{H}^{q_1}\partial_\beta\hat{H}\hat{H}^{q_2}]$
for pairs
$(q_1=p-2,q_2=R-p)$ for which there exists in the $\sum_p$ sum another pair
$(q'_1,q'_2)=(q_2,q_1)$ cancel against each other. 
In Eq.~\eqref{eq:40} we provide the full expression for the Berry curvature in
as a function of these terms.

\begin{widetext}

We find that for three-sublattice systems, dropping the band index
supercripts $(n)$ on the $\ell$'s and the $(\alpha\beta)$ superscripts
on the $\Lambda$'s to avoid clutter,
  \begin{align}
    \label{eq:40}
    &\Omega^{\alpha\beta}_{(n)}\nonumber\\
    & = 2{\rm Im}\Big[ -\Lambda _{(0,1)} \ell_1^3\nonumber\\
    &\quad\qquad\;\;-3   \ell_1^2\ell_2 \Lambda _{(0,2)}   \nonumber\\
    &\quad\qquad\;\;-3 \ell_1\ell_2^2   \Lambda _{(1,2)}-3\left(\ell_3\ell_1^2+
      \ell_2^2\ell_1\right) \Lambda_{(0,3)}\nonumber\\
    &\quad\qquad\;\;-\left(\ell_2^3+6\ell_1
      \ell_3\ell_2+3\ell_1^2\ell_4\right)\Lambda_{(0,4)}-2\left(\ell_2^3+3\ell_1   \ell_3 \ell_2\right)   \Lambda_{(1,3)}\nonumber\\
    &\quad\qquad\;\;-3\left(\ell_5 \ell_1^2+\ell_3^2   \ell_1+2\ell_2   \ell_4
      \ell_1+\ell_2^2 \ell_3\right)   \Lambda_{(0,5)}-3\left(2\ell_3 \ell_2^2+2\ell_1 \ell_4   \ell_2+\ell_1   \ell_3^2\right) \Lambda_{(1,4)}-3\left(\ell_3   \ell_2^2+ \ell_1   \ell_3^2\right) \Lambda_{(2,3)}\nonumber\\
    &\quad\qquad\;\;-3\left(\ell_4 \ell_2^2+\ell_3^2 \ell_2+2\ell_1 \ell_5
      \ell_2+2\ell_1   \ell_3 \ell_4\right)
      \Lambda_{(0,6)}-6\left(\ell_4   \ell_2^2+\ell_3^2   \ell_2+
      \ell_1   \ell_5 \ell_2+\ell_1 \ell_3   \ell_4\right)
      \Lambda_{(1,5)}\nonumber\\
    &\quad\qquad\;\;\qquad-3\left(\ell_4   \ell_2^2+2\ell_3^2   \ell_2+2\ell_1   \ell_3 \ell_4\right)   \Lambda_{(2,4)}\nonumber\\
    &\quad\qquad\;\;-\left(\ell_3^3+6   \ell_2 \ell_4   \ell_3+6 \ell_1   \ell_5
      \ell_3+3   \ell_1 \ell_4^2+3   \ell_2^2 \ell_5\right)
      \Lambda_{(0,7)}-\left(2   \ell_3^3+12 \ell_2   \ell_4 \ell_3+6
      \ell_1 \ell_5   \ell_3+3 \ell_1   \ell_4^2+6 \ell_2^2
      \ell_5\right) \Lambda_{(1,6)}\nonumber\\
    &\quad\qquad\;\;\qquad-3\left(\ell_3^3+4\ell_2   \ell_4 \ell_3+2\ell_1 \ell_5   \ell_3+\ell_1   \ell_4^2+\ell_2^2   \ell_5\right) \Lambda_{(2,5)}-\left(\ell_3^3+6   \ell_2 \ell_4   \ell_3+3 \ell_1   \ell_4^2\right) \Lambda_{(3,4)}\nonumber\\
    &\quad\qquad\;\;-3\left(\ell_4 \ell_3^2+2\ell_2 \ell_5   \ell_3+\ell_2
      \ell_4^2+2\ell_1   \ell_4 \ell_5\right)
      \Lambda_{(0,8)}-6\left(\ell_4   \ell_3^2+2\ell_2   \ell_5
      \ell_3+2\ell_2 \ell_4^2+2\ell_1 \ell_4   \ell_5\right)
      \Lambda_{(1,7)}\nonumber\\
    &\quad\qquad\;\;\qquad-3\left(3\ell_4   \ell_3^2+4\ell_2   \ell_5 \ell_3+2\ell_2 \ell_4^2+2\ell_1 \ell_4   \ell_5\right) \Lambda_{(2,6)}-6\left(\ell_4   \ell_3^2+\ell_2   \ell_5 \ell_3+\ell_2 \ell_4^2+\ell_1 \ell_4   \ell_5\right) \Lambda_{(3,5)}\nonumber\\
    &\quad\qquad\;\;-3\left(\ell_5 \ell_3^2+\ell_4^2 \ell_3+\ell_1
      \ell_5^2+2\ell_2 \ell_4   \ell_5\right)
      \Lambda_{(0,9)}-3\left(2 \ell_5   \ell_3^2+2 \ell_4^2   \ell_3+
      \ell_1   \ell_5^2+4 \ell_2   \ell_4 \ell_5\right)   \Lambda_{(1,8)}\nonumber\\
    &\quad\qquad\;\;\qquad-3\left(3 \ell_5   \ell_3^2+3\ell_4^2   \ell_3+ \ell_1
      \ell_5^2+4\ell_2   \ell_4 \ell_5\right)
      \Lambda_{(2,7)}-3\left(2\ell_5   \ell_3^2+3\ell_4^2   \ell_3+
      \ell_1   \ell_5^2+4 \ell_2   \ell_4 \ell_5\right)
      \Lambda_{(3,6)}\nonumber\\
    &\quad\qquad\;\;\qquad-3\left(\ell_5   \ell_3^2+ \ell_4^2   \ell_3+\ell_1   \ell_5^2+2\ell_2   \ell_4 \ell_5\right)   \Lambda_{(4,5)}\nonumber\\
    &\quad\qquad\;\;-\left(\ell_4^3+6   \ell_3 \ell_5   \ell_4+3 \ell_2
      \ell_5^2\right) \Lambda_{(0,10)}-2\left(\ell_4^3+6\ell_3
      \ell_5 \ell_4+3\ell_2 \ell_5^2\right)   \Lambda
      _{(1,9)}-3\left(\ell_4^3+6\ell_3   \ell_5 \ell_4+2
      \ell_2 \ell_5^2\right)   \Lambda_{(2,8)}\nonumber\\
    &\quad\qquad\;\;\qquad-2\left(2\ell_4^3+9\ell_3   \ell_5 \ell_4+3\ell_2 \ell_5^2\right)   \Lambda_{(3,7)}-2\left(\ell_4^3+6\ell_3   \ell_5 \ell_4+3\ell_2 \ell_5^2\right)   \Lambda_{(4,6)}\nonumber\\
    &\quad\qquad\;\;-3\left(\ell_4^2 \ell_5+\ell_3   \ell_5^2\right)
      \Lambda_{(0,11)}-6\left(\ell_4^2 \ell_5+\ell_3
      \ell_5^2\right)   \Lambda_{(1,10)}-9\left(\ell_5 \ell_4^2+
      \ell_3 \ell_5^2\right)   \Lambda _{(2,9)}-3\left(4\ell_5
      \ell_4^2+3\ell_3 \ell_5^2\right)   \Lambda_{(3,8)}\nonumber\\
    &\quad\qquad\;\;\qquad-9\left(\ell_5 \ell_4^2+\ell_3 \ell_5^2\right)\Lambda_{(4,7)}-3\left(  \ell_4^2 \ell_5+ \ell_3   \ell_5^2\right) \Lambda_{(5,6)}\nonumber\\
    &\quad\qquad\;\;-3 \ell_4   \ell_5^2 \Lambda_{(0,12)}-6 \ell_4   \ell_5^2 \Lambda_{(1,11)}-9 \ell_4   \ell_5^2 \Lambda _{(2,10)}-12 \ell_4   \ell_5^2 \Lambda_{(3,9)}-12 \ell_4   \ell_5^2 \Lambda_{(4,8)}-6 \ell_4   \ell_5^2 \Lambda _{(5,7)}\nonumber\\
    &\quad\qquad\;\;-\ell_5^3 \Lambda_{(0,13)}-2   \ell_5^3 \Lambda_{(1,12)}-3   \ell_5^3 \Lambda_{(2,11)}-4   \ell_5^3 \Lambda_{(3,10)}-5\ell_5^3 \Lambda_{(4,9)}-3   \ell_5^3 \Lambda_{(5,8)}-\ell_5^3 \Lambda   _{(6,7)}
     \Big].
  \end{align}
\end{widetext}
coefficients in front of $\chi_{123}$ for the kagom\'e lattice

On the spin-orbit-coupling-free kagom\'e lattice, for
$\alpha\beta=xy$, and defining $\Lambda_{(q_1,q_2)}^{{\rm im}}={\rm Im}\Lambda_{(q_1,q_2)}$,
\begin{align}
  \label{eq:17}
  &\Lambda_{(0,1)}^{{\rm im}}=\Lambda_{(0,2)}^{{\rm im}}=\Lambda_{(0,3)}^{{\rm im}}=\Lambda_{(1,2)}^{{\rm im}}=\Lambda_{(0,4)}^{{\rm im}}=0,\nonumber\\
  &\Lambda_{(1,3)}^{{\rm im}}\nonumber\\
  &\;=2\sqrt{3}K^3(3-\cos^2k_x+\sin^2k_x-2\cos k_x\cos(\sqrt{3}k_y))\chi_{123}.
\end{align}

\section{Three-sublattice triangular lattice}
\label{sec:triangular}

\subsection{Explicit form of the Hamiltonian}
\label{sec:expl-form-hamilt}

\begin{widetext}
On the triangular lattice (with $C_3$ symmetry)
\begin{align}
  \label{eq:52}
  \hat{H}(\bs{k})=&\begin{pmatrix}
  \vec{S}_1\cdot\vec{\sigma} &
  t_0(e^{i\bs{k}\cdot\bs{e}_{12(1)}}+e^{i\bs{k}\cdot\bs{e}_{12(2)}}+e^{i\bs{k}\cdot\bs{e}_{12(3)}})
  &
  t_0(e^{i\bs{k}\cdot\bs{e}_{13(1)}}+e^{i\bs{k}\cdot\bs{e}_{13(2)}}+e^{i\bs{k}\cdot\bs{e}_{13(3)}})
  \\
  t_0(e^{i\bs{k}\cdot\bs{e}_{21(1)}}+e^{i\bs{k}\cdot\bs{e}_{21(2)}}+e^{i\bs{k}\cdot\bs{e}_{21(3)}})
  & \vec{S}_2\cdot\vec{\sigma} &
  t_0(e^{i\bs{k}\cdot\bs{e}_{23(1)}}+e^{i\bs{k}\cdot\bs{e}_{23(2)}}+e^{i\bs{k}\cdot\bs{e}_{23(3)}})\\
  t_0(e^{i\bs{k}\cdot\bs{e}_{31(1)}}+e^{i\bs{k}\cdot\bs{e}_{31(2)}}+e^{i\bs{k}\cdot\bs{e}_{31(3)}})
  & t_0(e^{i\bs{k}\cdot\bs{e}_{32(1)}}+e^{i\bs{k}\cdot\bs{e}_{32(2)}}+e^{i\bs{k}\cdot\bs{e}_{32(3)}})
  & \vec{S}_3\cdot\vec{\sigma}
\end{pmatrix},
\end{align}
\end{widetext}
and so 
\begin{align}
 \hat{H}(\bs{k}) &=\begin{pmatrix}
  \vec{S}_1\cdot\vec{\sigma} &
  t_0F(\bs{k})
  &
  t_0^*F^*(\bs{k})
  \\
  t_0^*F^*(\bs{k})
  & \vec{S}_2\cdot\vec{\sigma} &
  t_0F(\bs{k})\\
  t_0F(\bs{k})
  & t_0^*F^*(\bs{k})
  & \vec{S}_3\cdot\vec{\sigma}
\end{pmatrix},
\end{align}
where
\begin{equation}
  \label{eq:53}
  F(\bs{k})=e^{i\bs{k}\cdot\bs{e}_{12}}+e^{i\bs{k}\cdot\bs{e}_{23}}+e^{i\bs{k}\cdot\bs{e}_{31}}.
\end{equation}
We also have
\begin{align}
  \label{eq:54}
  \partial_\gamma\hat{H}(\bs{k}) &=\begin{pmatrix}
  0 &
  it_0I^\gamma(\bs{k})
  &
  -it_0^*(I^\gamma)^*(\bs{k})
  \\
  -it_0^*(I^\gamma)^*(\bs{k})
  & 0 &
  it_0I^\gamma (\bs{k})\\
  it_0I^\gamma (\bs{k})
  & -it_0^*(I^\gamma)^* (\bs{k})
  & 0
\end{pmatrix},
\end{align}
where
\begin{equation}
  \label{eq:55}
  I^\gamma(\bs{k})=-i\partial_\gamma F(\bs{k})={e}_{12}^\gamma e^{i\bs{k}\cdot\bs{e}_{12}}+{e}_{23}^\gamma e^{i\bs{k}\cdot\bs{e}_{23}}+{e}_{31}^\gamma e^{i\bs{k}\cdot\bs{e}_{31}}.
\end{equation}

\subsection{Vanishing Berry curvature}
\label{sec:vanish-berry-curv}

Here we consider Eq.~\eqref{eq:503} and apply it to the case of a
three-sublattice triangular lattice (where the unit cell is a triangle
of nearest-neighbors) where, for $a\neq b$
\begin{equation}
  \label{eq:49}
  {\sf
  h}^\mu_{ab,(\eta)}=t_0^\mu e^{i\bs{k}\cdot\bs{e}_{ab(\eta)}},
\end{equation}
i.e.\ $t_{ab,(\eta)}^\mu$ is actually independent of $a,b$ and
$\eta$. This
is in particular the case in the absence of spin-orbit coupling,
where, additionally we have $t_0^\mu=t_0\delta_{\mu0}$. Then,
Eq.~\eqref{eq:503} becomes (recall we defined
$\bs{I}_{ab}\equiv\sum_\eta\bs{e}_{ab(\eta)}e^{i\bs{k}\cdot{e}_{ab(\eta)}}$,
and $\bs{I}\equiv\bs{I}_{12}$)
\begin{align}
  \label{eq:50}
  &J_{a_1,a_2,a_p,a_{p+1}}^{\alpha\beta|\mu_1\mu_p}\\
  &=-t_0^{\mu_1}t_0^{\mu_p}\sum_{\eta_1,\eta_p}e_{a_1a_2(\eta_1)}^\alpha
  e_{a_pa_{p+1}(\eta_p)}^\beta
    e^{i\bs{k}\cdot(\bs{e}_{a_1a_2(\eta_1)}+\bs{e}_{a_pa_{p+1}(\eta_p)})}\nonumber\\
  &=-t_0^{\mu_1}t_0^{\mu_p}I_{a_1a_2}^\alpha
  I_{a_pa_{p+1}}^\beta,\nonumber
\end{align}


As
mentioned in the main text, we have
\begin{equation}
  \label{eq:8}
 \bs{I}\equiv\bs{I}^{\rm t}_{12}=\bs{I}^{\rm t}_{23}=\bs{I}^{\rm t}_{31}=-(\bs{I}^{\rm t}_{21})^*=-(\bs{I}^{\rm t}_{32})^*=-(\bs{I}^{\rm t}_{13})^*,
\end{equation}
where $\bs{I}^{\rm t}_{ab}$ is
defined in Eq.~\eqref{eq:505}. Now, under
$\alpha\leftrightarrow\beta$, we have
\begin{align}
  \label{eq:58}
  I^\alpha_{a_1a_2}I^\beta_{a_pa_{p+1}}&\rightarrow
                                         I^\beta_{a_1a_2}I^\alpha_{a_pa_{p+1}}\\
  &\rightarrow(I^\beta_{a_pa_{p+1}}I^\alpha_{a_1a_{2}})^{\upsilon_{a_1a_2}\upsilon_{a_pa_{p+1}}},\nonumber
\end{align}
where we used Eq.~\eqref{eq:8} in the second line, and where, when we use $\upsilon_{ab}$ as an exponent, we mean
$A^{\upsilon_{ab}}=A^*$ for $\upsilon_{ab}=-1$ and $A^{\upsilon_{ab}}=A$ for
$\upsilon_{ab}=1$. We may also write
\begin{align}
  \label{eq:59}
  I^\alpha_{a_1a_2}I^\beta_{a_pa_{p+1}}=\upsilon_{a_1a_2}\upsilon_{a_pa_{p+1}} (I^\alpha)^{\upsilon_{a_1a_{2}}} (I^\beta)^{\upsilon_{a_pa_{p+1}}}.
\end{align}
We can distinguish two cases.
\begin{enumerate}
\item The first is that where
$\upsilon_{a_1a_2}\upsilon_{a_pa_{p+1}}=1$---which means that both
$a_1a_2$ and $a_pa_{p+1}$ are ``ordered'' or both are ``disordered''---in which case we have immediately
\begin{equation}
\label{eq:9}
I^\alpha_{a_1a_2}I^\beta_{a_pa_{p+1}} \rightarrow I^\beta_{a_1a_2}I^\alpha_{a_pa_{p+1}}=I^\alpha_{a_1a_2}I^\beta_{a_pa_{p+1}},
\end{equation}
(note however that the above is equal to $I^\alpha I^\beta$ {\em or} to $(I^\alpha I^\beta)^*$)
and so those contribute
$\left.\Omega^{\beta\alpha}\right|_1=\left.\Omega^{\alpha\beta}\right|_1$,
and so zero.
\item In the second
case, where $\upsilon_{a_1a_2}\upsilon_{a_pa_{p+1}}=-1$, we have,
under $\alpha\leftrightarrow\beta$
\begin{equation}
 \label{eq:10}
I^\alpha_{a_1a_2}I^\beta_{a_pa_{p+1}} \rightarrow I^\beta_{a_1a_2}I^\alpha_{a_pa_{p+1}} =(I^\alpha_{a_1a_2}I^\beta_{a_pa_{p+1}})^*. 
\end{equation}
We can rewrite the above as
\begin{equation}
  \label{eq:60}
  -I^\alpha(I^\beta)^* \rightarrow -I^\beta (I^\alpha)^* =(I^\alpha (-I^\beta)^*)^*
\end{equation}
or as
\begin{equation}
  \label{eq:61}
  -(I^\alpha)^*I^\beta \rightarrow -(I^\beta)^* I^\alpha =((-I^\alpha)^* I^\beta)^*.
\end{equation}
The sum of those terms in ${\rm Tr}[\Lambda_{(q_1,q_2)}^{\alpha\beta}]$ which
fall in case 2 are
\begin{align}
  \label{eq:64}
  &\left.{\sf Tr}[\Lambda_{(q_1,q_2)}^{\alpha\beta}]\right|_{-1}=\nonumber\\
  &|t_0|^2 I^\alpha
    (I^\beta)^*\nonumber\\
  &{\sf Tr}[(\hat{H}^{q_1})_{21}(\hat{H}^{q_2})_{31}+(\hat{H}^{q_1})_{12}(\hat{H}^{q_2})_{13}\nonumber\\
  &+(\hat{H}^{q_1})_{23}(\hat{H}^{q_2})_{21}+(\hat{H}^{q_1})_{32}(\hat{H}^{q_2})_{12}\nonumber\\
  &+(\hat{H}^{q_1})_{31}(\hat{H}^{q_2})_{32}+(\hat{H}^{q_1})_{13}(\hat{H}^{q_2})_{23}\nonumber\\
                     &+(\hat{H}^{q_1})_{11}(\hat{H}^{q_2})_{33}+(\hat{H}^{q_1})_{22}(\hat{H}^{q_2})_{11}+(\hat{H}^{q_1})_{33}(\hat{H}^{q_2})_{22}]\nonumber\\
&  +|t_0|^2 (I^\alpha)^*
 I^\beta{\sf Tr}[(\hat{H}^{q_1})_{31}(\hat{H}^{q_2})_{21}+(\hat{H}^{q_1})_{13}(\hat{H}^{q_2})_{12}\nonumber\\
  &+(\hat{H}^{q_1})_{21}(\hat{H}^{q_2})_{23}+(\hat{H}^{q_1})_{12}(\hat{H}^{q_2})_{32}\nonumber\\
  &+(\hat{H}^{q_1})_{32}(\hat{H}^{q_2})_{31}+(\hat{H}^{q_1})_{23}(\hat{H}^{q_2})_{13}\nonumber\\
  &+(\hat{H}^{q_1})_{33}(\hat{H}^{q_2})_{11}+(\hat{H}^{q_1})_{11}(\hat{H}^{q_2})_{22}+(\hat{H}^{q_1})_{22}(\hat{H}^{q_2})_{33}],
\end{align}
where ${\sf Tr}$ is a Pauli-space ($2\times2$) trace and $\hat{\sf
    H}_{\mu_1}\cdots\hat{\sf H}_{\mu_{q_1}}$ are sublattice matrix
  ($3\times3$) multiplications. 

We can show explicitly that
\begin{align}
   & {\sf Tr}[(\hat{H}^{q_1})_{21}(\hat{H}^{q_2})_{31}+(\hat{H}^{q_1})_{12}(\hat{H}^{q_2})_{13}\nonumber\\
  &\quad+(\hat{H}^{q_1})_{23}(\hat{H}^{q_2})_{21}+(\hat{H}^{q_1})_{32}(\hat{H}^{q_2})_{12}\nonumber\\
    &\quad+(\hat{H}^{q_1})_{31}(\hat{H}^{q_2})_{32}+(\hat{H}^{q_1})_{13}(\hat{H}^{q_2})_{23}]\nonumber\\
      &={\sf Tr}[(\hat{H}^{q_1})_{31}(\hat{H}^{q_2})_{21}+(\hat{H}^{q_1})_{13}(\hat{H}^{q_2})_{12}\nonumber\\
  &\quad+(\hat{H}^{q_1})_{21}(\hat{H}^{q_2})_{23}+(\hat{H}^{q_1})_{12}(\hat{H}^{q_2})_{32}\nonumber\\
  &\quad+(\hat{H}^{q_1})_{32}(\hat{H}^{q_2})_{31}+(\hat{H}^{q_1})_{23}(\hat{H}^{q_2})_{13}]\in\mathbb{R},
\end{align}
and
\begin{align}
    \label{eq:7}
  &  {\sf
  Tr}[(\hat{H}^{q_1})_{11}(\hat{H}^{q_2})_{33}+(\hat{H}^{q_1})_{22}(\hat{H}^{q_2})_{11}+(\hat{H}^{q_1})_{33}(\hat{H}^{q_2})_{22}]\nonumber\\
  &={\sf
  Tr}[ (\hat{H}^{q_1})_{33}(\hat{H}^{q_2})_{11}+(\hat{H}^{q_1})_{11}(\hat{H}^{q_2})_{22}+(\hat{H}^{q_1})_{22}(\hat{H}^{q_2})_{33}]\in\mathbb{R},
\end{align}
and so $\left.{\sf
    Tr}[\Lambda_{(q_1,q_2)}^{\alpha\beta}]\right|_{-1}\in\mathbb{R}$,
so that it does not contribute to the Berry curvature.
\end{enumerate}
Combining cases 1 and 2 together, we have shown that the Berry curvature
for the spin-orbit-coupling-free triangular lattice vanishes,
regardless of the spin configuration.

\end{document}